\documentclass[aps,prl,twocolumn,superscriptaddress,floatfix,notitlepage]{revtex4-2}

\usepackage{hyperref}
\usepackage{fullpage}
\usepackage{comment}
\usepackage{lineno}
\usepackage{graphicx,xcolor}
\usepackage{amssymb}
\usepackage{braket}
\usepackage[frozencache]{minted}
\usepackage{physics}
\usepackage{yquant}
\usepackage{tikz}
\usepackage[caption=false]{subfig}
\usepackage{orcidlink}
\usetikzlibrary{quantikz,fit,quotes,svg.path}

\newcommand{\h}[1]{\mintinline{haskell}{#1}}
\newcommand{\x}{\textsc{x}}
\newcommand{\cx}{\textsc{cx}}
\newcommand{\ccx}{\textsc{ccx}}
\newcommand{\cccx}{\textsc{cccx}}

\newcommand{\preim}[2]{\{\cdot\stackrel{#1}{\longleftarrow}{#2}\}}
\newcommand{\finset}[1]{[\mathbf{#1}]}
\newcommand{\red}[1]{{\color{red}{#1}}}
\newcommand{\blue}[1]{{\color{blue}{#1}}}

\begin{document}

\title{Retrodictive Quantum Computing}

\begin{abstract}
Quantum models of computation are widely believed to be more powerful
than classical ones. Efforts center on proving that, for a given
problem, quantum algorithms are more resource efficient than any
classical one. All this, however, assumes a standard predictive
paradigm of reasoning where, given initial conditions, the future
holds the answer. How about bringing information from the future to
the present and exploit it to one's advantage? This is a radical new
approach for reasoning, so-called Retrodictive Computation, that
benefits from the specific form of the computed functions. We
demonstrate how to use tools of symbolic computation to realize
retrodictive quantum computing at scale and exploit it to efficiently,
and classically, solve instances of the quantum Deutsch-Jozsa,
Bernstein-Vazirani, Simon, Grover, and Shor's algorithms.
\end{abstract}

\author{Jacques Carette \orcidlink{0000-0001-8993-9804}}
\affiliation{\mbox{Department of Computer Science, McMaster University, Hamilton, Ontario L8S 4K1, Canada}}
\author{Gerardo Ortiz \orcidlink{0000-0003-3254-4494}}
\email{ortizg@iu.edu}
\affiliation{\mbox{Department of Physics, Indiana University, Bloomington, Indiana 47405, USA}}
\author{Amr Sabry \orcidlink{0000-0002-1025-7331}}
\affiliation{\mbox{Department of Computer Science, Indiana University, Bloomington, Indiana 47405, USA}}

\maketitle

%%%%%%%%%%%%%%%%%%%%%%%%%%%%%%%%%%%%%%%%%%%%%%%%%%%%%%%%%%%%%%%%%%%%%%%%%%%%%%%%%%%%%%%%%%

{\it Introduction.--} Quantum evolution is time-reversible and yet this reversibility 
is not fully exploited in the circuit model of quantum
computing. Indeed, most quantum algorithms expressed in the circuit
model compute strictly from the present to the future, preparing
initial states and proceeding forward with unitary transformations and
measurements. We may call this predictive computation. In contrast, retrodictive quantum
theory~\cite{sym13040586}, retrocausality~\cite{Aharonov2008}, and the
time-symmetry of physical laws~\cite{RevModPhys.27.179} all suggest
that quantum computation embodies richer --untapped-- modes of
computation, which could exploit knowledge about the future for a
computational advantage.

We demonstrate that by using symbolic partial
evaluation~\cite{futamura}, retrodictive reasoning can indeed be used
as a computational resource that exhibits richer modes of computation
at the boundary of the classical/quantum divide. Specifically, instead
of fully specifying the initial conditions of a quantum circuit and
computing forward, it is possible to compute, classically, in both the
forward and backward directions starting from partially-specified
initial and final conditions. Furthermore, this mixed mode of
computation (i) can solve problems with fewer resources than the
conventional forward mode of execution, sometimes even purely
classically (de-quantization), (ii) can be expressed in a symbolic
representation that immediately exposes global relational properties
of the wavefunction that are needed for quantum algorithms, and (iii)
reveals that the entanglement patterns inherent in genuine quantum
algorithms with no known classical counterparts are artifacts of the
chosen symbolic representation.

\begin{figure}[b]
\subfloat[Bell circuit\label{fig:bell}]{%
    \begin{tikzpicture}[scale=0.80]
   \begin{yquant*}[register/minimum height=0.9cm]
   qubit {$\ket0$} x;
   qubit {$\ket0$} y;
   box {$H$} x;
   cnot y | x;
   measure x;
   measure y;
  \end{yquant*}
\end{tikzpicture}
}
\quad
\subfloat[Quantum core\label{fig:bellqcore}]{%
    \begin{tikzpicture}[scale=0.90]
   \begin{yquant*}[register/minimum height=0.9cm]
   qubit {$\ket{x_1}$} x;
   qubit {$\ket{y_1}$} y;
   box {$H$} x;
   cnot y | x;
   output {$\ket{x_2}$} x;
   output {$\ket{y_2}$} y;
  \end{yquant*}
\end{tikzpicture}
}
\quad
\subfloat[Classical core\label{fig:bellccore}]{%
    \begin{tikzpicture}[scale=0.9]
   \begin{yquant*}[register/minimum height=0.9cm]
   qubit {$\ket{x_1}$} x;
   qubit {$\ket{y_1}$} y;
   hspace {0.1cm} -;   
   cnot y | x;
   hspace {0.1cm} -;   
   output {$\ket{x_2}$} x;
   output {$\ket{y_2}$} y;
  \end{yquant*}
\end{tikzpicture}
}
\caption{\label{fig:bellall} (a) A conventional quantum circuit with
  initial conditions and measurement; (b) its quantum core without
  measurement and with unspecified initial and final conditions; and
  (c) its classical core without explicit quantum superpositions.}
\end{figure}
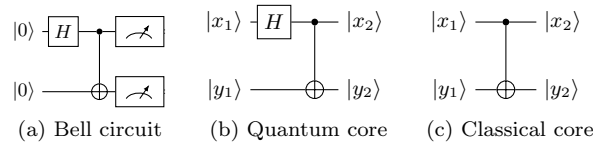

The main ideas underlying our contributions can be illustrated with
the aid of the small examples in Fig.~\ref{fig:bellall}. In the
conventional computational predictive mode (Fig.~\ref{fig:bell}), the
execution starts with the initial state $\ket{00}$. The first gate
(Hadamard) evolves the state to $1/\sqrt{2}(\ket{00} + \ket{10})$
which is transformed by the controlled-not (\cx) gate to
$1/\sqrt{2}(\ket{00} + \ket{11})$
(\hyperref[sec:Methods]{Appendix}). The measurements at the end
produce 00 or 11 with equal probability. Fig.~\ref{fig:bellqcore}
keeps the quantum core of the circuit, removing the measurements, and
naming the inputs and outputs with symbolic variables. Now, instead of
setting initial values $x_1=y_1=0$ and computing forward as before, we
can, for example, set final values $x_2=1$ and $y_2=0$ and calculate
backwards as follows: $\ket{10}$ evolves in the backwards direction to
$\ket{11}$ and then to $1/\sqrt{2}(\ket{01}-\ket{11})$. In other
words, in order to observe $x_2y_2=10$, the variable $x_1$ should be
prepared in the superposition $1/\sqrt{2}(\ket{0}-\ket{1})$ and $y_1$
should be prepared in the state $\ket{1}$. More interestingly, we can
partially specify the initial and final conditions. For example, we
can fix $x_1=0$ and $x_2=1$ and ask if there are any possible values
for~$y_1$ and $y_2$ that would be consistent with this setting.  Using
the techniques of symbolic partial evaluation
(\hyperref[sec:Methods]{Appendix}), we calculate as follows. The
initial state is $\ket{0y_1}$ which evolves to
$1/\sqrt{2}(\ket{0y_1}+\ket{1y_1})$ and then to
$1/\sqrt{2}(\ket{0y_1}+\ket{1(1 \oplus y_1)})$ where $\oplus$ is the
exclusive-or operation and $1 \oplus y_1$ is the canonical way of
negating~$y_1$ in the Algebraic Normal Form (ANF) of boolean
expressions (\hyperref[sec:Methods]{Appendix}).  This final state can
now be reconciled with the specified final conditions $1y_2$ revealing
that the settings are consistent provided that $y_2 = 1 \oplus
y_1$. Had we additionally imposed $y_1=y_2$, the equation is
unsatisfiable. We can, in fact, go one step further and analyze the
circuit without the Hadamard gate as shown in
Fig.~\ref{fig:bellccore}. The reasoning is that the role of Hadamard
is to introduce (modulo a phase) uncertainty about whether $x_1=0$ or
$x_1=1$. But, again modulo a phase, the same uncertainty can be
expressed by just using the variable~$x_1$. Thus, in
Fig.~\ref{fig:bellccore}, we can set $y_1=0$ and $y_2=1$ and ask about
values of $x_1$ and~$x_2$ that would be consistent with this
setting. Calculating backwards from $\ket{x_21}$, the state evolves to
$\ket{x_2(1 \oplus x_2)}$ which can be reconciled with the initial
conditions yielding the constraints $x_1=x_2$ and $1 \oplus x_2 = 0$
whose solutions are $x_1 = x_2 = 1$. Finally, we can leave $x_1$ as a
symbolic variable, set $y_1=0$, and evaluate the circuit starting with
the state $\ket{x_10}$. The result, $\ket{x_1x_1}$, calculated in a
single step, expresses the entanglement relation between the two
qubits by using the same symbol twice. Classically, this correlation
would need multiple evaluations to be inferred. Quantum mechanically,
this correlation would not be directly observable. In that sense,
symbolic execution is a new mode of execution which has distinct
advantages in some situations.

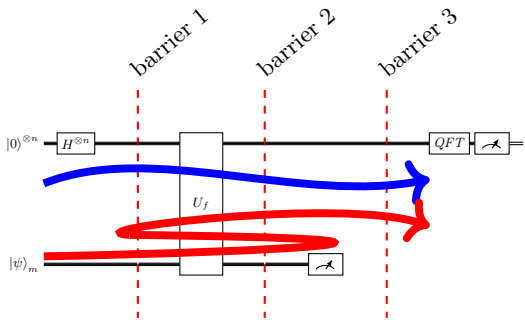
\begin{figure}[b]
  \begin{tikzpicture}[scale=0.57,every label/.style={rotate=40, anchor=south west}]
    \begin{yquant*}[operators/every barrier/.append style={red, thick}
        , operator/minimum width=3mm
        , operator/separation=5mm
        , register/separation=25mm
        ]
    qubits {$\ket0^{\otimes n}$} a;
    qubits {$\ket{\psi}_m$} b;
    box {$H^{\otimes n}$} a;
    ["barrier 1"]
    barrier (-);
    [x radius=5mm, y radius=5mm]
    box {$U_f$} (a,b);
    ["barrier 2"]
    barrier (-);
    measure b;
    discard b;
    ["barrier 3"]
    barrier (-);
    box {$\mathit{QFT}$} a;
    measure a;
    \end{yquant*}
    \draw[line width=3pt, ->, blue] (0,-1.1) .. controls (2.5,-0.1) and (6,-1.6) .. (9,-1);
  \draw[line width=3pt, red] (0,-2.8) .. controls (5.5,-2.7) and (11,-2.4) .. (2,-2.3);
  \draw[line width=3pt, ->, red] (2,-2.3) .. controls (0.5,-2.2) and (7,-1.4) .. (9,-2.1);
  \end{tikzpicture}
\caption{\label{fig:templateQC}Template quantum circuit with flow of information indicated by arrows: Conventional (blue) and retrodictive flows (red). QFT: quantum Fourier transform.}
\end{figure}

These insights are robust and can be implemented in software
(\hyperref[sec:Methods]{Appendix}) to analyze circuits with millions
of gates for the quantum algorithms that match the template in
Fig.~\ref{fig:templateQC} (including Deutsch, Deutsch-Jozsa,
Bernstein-Vazirani, Simon, Grover, and Shor's
algorithms~\cite{doi:10.1137/S0097539796300921,deutsch,deutschJozsa,365701,doi:10.1137/S0097539795293172,mermin_2007,nielsen_chuang_2010,10.1145/237814.237866}). The
software is completely classical, performing mixed mode executions of
the classical core of the circuits, i.e., the $U_f$ block formally
defined as $U_f(\ket{x}\ket{y}) = \ket{x}\ket{f(x) \oplus
  y}$. Specifically, in all these algorithms, there is a top
collection of wires (which, following the standard predictive flow of
information, we call the input register) and a bottom collection of
wires (the output register). The input register is prepared in a
uniform superposition which can be represented using symbolic
variables. The measurement of the output register after barrier 2
provides partial information about the future which is, together with
the initial conditions of the output register, sufficient to
symbolically execute the circuit. In each case, instead of the
conventional execution flow depicted in Fig.~\ref{fig:templateQC}, we
find a possible measurement outcome $w$ at barrier (2) and perform a
symbolic retrodictive execution with a state $\ket{xw}$ going
backwards to collect the constraints on $x$ that enable us to solve
the problem in question (Fig.~\ref{fig:templateQC}).

{\it Algorithms.--} 
The accompanying code includes retrodictive implementations of six
major quantum algorithms: Deutsch, Deutsch-Jozsa, Bernstein-Vazirani,
Simon, Grover, and Shor \cite{mermin_2007,nielsen_chuang_2010} 
(\hyperref[sec:Methods]{Appendix}). 
An important insight is that each of these algorithms is essentially asking
the same question: \emph{which input states can produce a particular output
measurement?} In conventional predictive quantum execution, the 
question is answered by initializing the input register to a superposition of all
possible input states, evolving them through the $U_f$ block, and measuring
the output causing a phase kickback effect to refine the states to the ones
consistent with the measurement. A conceptually simpler solution is to just
start a retrodictive execution with the output measurement as a valid ``retrodictive 
input."

A word of
caution: all the algorithms (except Shor) are conventionally 
presented in the ``black-box'' complexity model where the internals of the $U_f$ 
are unspecified, where each access to $U_f$ counts
as one unit of execution cost, and where the algorithm complexity is expressed
using the number of times $U_f$ is queried. This is not, however, the only way to
analyze complexity~\cite{10.1145/3341106}. In our case, the $U_f$ block is an actual 
known circuit that we execute (usually symbolically). We return to this point
after highlighting the salient results for the
first five algorithms, and discussing the interesting case of
Shor's algorithm in detail.

\begin{figure}[t]
\centering
\begin{tikzpicture}[scale=1.2]
   \begin{yquant*}[register/minimum height=0.8cm]
   qubit {$\ket0$} x;
   qubit {$\ket1$} y;
   box {$H$} x;
   box {$H$} y;
   box {$U_f$} (x,y);
   measure y;
   align -;
   discard y;
   box {$H$} x;
   measure x;
  \end{yquant*}
\end{tikzpicture}
\caption{\label{fig:deutsch}Quantum Circuit for the Deutsch-Jozsa
  Algorithm $(n=1)$}
\end{figure}
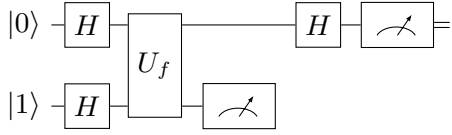

{\it De-Quantization.--} We abbreviate the set $\{ 0,1,\ldots,(n-1)\}$ as~$\finset{n}$. In the
Deutsch-Jozsa problem, we are given a function $\finset{2^\mathit{n}}
\rightarrow \finset{2}$ that is promised to be constant or balanced
and we need to distinguish the two cases. The quantum circuit
Fig.~\ref{fig:deutsch} shows the algorithm for the case $n=1$. Instead
of the conventional execution, we perform a retrodictive execution of
the $U_f$ block with an output measurement~$0$, i.e., with the state
$\ket{x_{n-1}\cdots x_1x_00}$.  The result of the execution is a
symbolic formula that determines the conditions under which
$f(x_{n-1},\cdots,x_0) = 0$. When the function is constant (say $0$), the
results are $0=0$ (always) or $1=0$ (never). When the function is
balanced, we get a formula that mentions the relevant variables. Thus, retrodictive 
reasoning does de-quantize Deutsch-Jozsa's problem for an arbitrary $\finset{2^\mathit{n}}
\rightarrow \finset{2}$ boolean function. For
example, here are the results for three different balanced
functions $\finset{2^\mathrm{6}} \rightarrow \finset{2}$:
\[\begin{array}{ll}
\textrm{Case 1.} & x_0 = 0 \\
\textrm{Case 2.} & x_0 \oplus x_1 \oplus x_2 \oplus x_3 \oplus
    x_4 \oplus x_5 = 0 \\
\textrm{Case 3.} & 1 \oplus x_3x_5 \oplus x_2x_4 \oplus x_1x_5
\oplus x_0x_3 \oplus x_0x_2 \oplus \\
& x_3x_4x_5 \oplus x_2x_3x_5 \oplus x_1x_3x_5 \oplus \\
& x_0x_3x_5 \oplus x_0x_1x_4 \oplus x_0x_1x_2 \oplus \\
& x_2x_3x_4x_5 \oplus x_1x_3x_4x_5 \oplus x_1x_2x_4x_5 \oplus \\
& x_1x_2x_3x_5 \oplus x_0x_3x_4x_5 \oplus x_0x_2x_4x_5 \oplus \\
& x_0x_2x_3x_5 \oplus x_0x_1x_4x_5 \oplus x_0x_1x_3x_5 \oplus \\
& x_0x_1x_3x_4 \oplus x_0x_1x_2x_4 \oplus x_0x_1x_2x_4x_5 \oplus \\
& x_0x_1x_2x_3x_5 \oplus x_0x_1x_2x_3x_4 = 0
\end{array}\]
In the first case, the function is balanced because it produces $0$
exactly when $x_0=0$ which happens half of the time in all possible
inputs; in the second case the output of the function is the
exclusive-or of all the input variables which is another easy instance
of a balanced function. The last case is a cryptographically-strong
function whose output pattern is balanced but, by design,
difficult to discern~\cite{quteprints21763}. An important insight is that we actually do not care about the exact
formula. Indeed, since we are promised that the function is either
constant or balanced, then any formula that refers to at least one
variable must indicate a balanced function: the
outcome of the algorithm can be immediately decided if the formula is
anything other than~0 or~1. Indeed, our implementation correctly
identifies all 12,870 balanced functions $\finset{2^\mathrm{4}} \rightarrow
\finset{2}$. This is significant as some of these functions produce
complicated entangled patterns during quantum evolution and could not
be de-quantized using previous approaches~\cite{djdeq}.

\begin{figure}[!]
\begin{tabular}{ll}
$u=0$ & 
  $\red{1} \oplus x_3 \oplus x_2 \oplus x_1 \oplus x_0 \oplus x_2x_3 \oplus x_1x_3 \oplus x_1x_2 \oplus$ \\
  &\quad $x_0x_3 \oplus x_0x_2 \oplus x_0x_1 \oplus x_1x_2x_3 \oplus x_0x_2x_3 \oplus$ \\
  &\quad $x_0x_1x_3 \oplus x_0x_1x_2 \oplus x_0x_1x_2x_3$ \\
$u=1$ & 
  $\red{x_0} \oplus x_0x_3 \oplus x_0x_2 \oplus x_0x_1 \oplus x_0x_2x_3 \oplus x_0x_1x_3 \oplus$ \\
  &\quad $x_0x_1x_2 \oplus x_0x_1x_2x_3$ \\
$u=2$ &
  $\red{x_1} \oplus x_1x_3 \oplus x_1x_2 \oplus x_0x_1 \oplus x_1x_2x_3 \oplus x_0x_1x_3 \oplus$ \\
  &\quad $x_0x_1x_2 \oplus x_0x_1x_2x_3$ \\
$u=3$ &
  $\red{x_0x_1} \oplus x_0x_1x_3 \oplus x_0x_1x_2 \oplus x_0x_1x_2x_3$ \\
$u=4$ &
  $\red{x_2} \oplus x_2x_3 \oplus x_1x_2 \oplus x_0x_2 \oplus x_1x_2x_3 \oplus x_0x_2x_3 \oplus$ \\
  &\quad $x_0x_1x_2 \oplus x_0x_1x_2x_3$ \\
$u=5$ &
  $\red{x_0x_2} \oplus x_0x_2x_3 \oplus x_0x_1x_2 \oplus x_0x_1x_2x_3$ \\
$u=6$ &
  $\red{x_1x_2} \oplus x_1x_2x_3 \oplus x_0x_1x_2 \oplus x_0x_1x_2x_3$ \\
$u=7$ &
  $\red{x_0x_1x_2} \oplus x_0x_1x_2x_3$ \\
$u=8$ &
  $\red{x_3} \oplus x_2x_3 \oplus x_1x_3 \oplus x_0x_3 \oplus x_1x_2x_3 \oplus x_0x_2x_3 \oplus$ \\
  &\quad $x_0x_1x_3 \oplus x_0x_1x_2x_3$ \\
$u=9$ &
  $\red{x_0x_3} \oplus x_0x_2x_3 \oplus x_0x_1x_3 \oplus x_0x_1x_2x_3$ \\
$u=10$ &
  $\red{x_1x_3} \oplus x_1x_2x_3 \oplus x_0x_1x_3 \oplus x_0x_1x_2x_3$ \\
$u=11$ &
  $\red{x_0x_1x_3} \oplus x_0x_1x_2x_3$ \\
$u=12$ &
  $\red{x_2x_3} \oplus x_1x_2x_3 \oplus x_0x_2x_3 \oplus x_0x_1x_2x_3$ \\
$u=13$ &
  $\red{x_0x_2x_3} \oplus x_0x_1x_2x_3$ \\
$u=14$ &
  $\red{x_1x_2x_3} \oplus x_0x_1x_2x_3$ \\
$u=15$ &
  $\red{x_0x_1x_2x_3}$
\end{tabular}
\caption{\label{fig:Grover}Result of retrodictive execution for the Grover oracle ($n=4$, $u$ in the range $\{0,\cdots,15\}$). The highlighted red subformula is the binary representation of the hidden input $u$.}
\end{figure}
The above analysis suggests that the specific equations may
not be relevant for some algorithms and that it is enough to glean the equations
to solve the problem in question. Indeed, this observation holds for not just 
the Deutsch-Jozsa algorithm but also for the 
Bernstein-Vazirani, Simon, and Grover algorithms. In all cases,
the result can be immediately read from the formula sidestepping
the underlying $\mathit{NP}$-complete SAT
problem~\cite{4640789,Karp1972,10.1145/800157.805047}. In the
Bernstein-Vazirani case, formulae are guaranteed to have all subformulae consist of single variables, e.g., 
$x_1 \oplus x_3 \oplus x_4 \oplus x_5$; the secret string is then the
binary number that has a 1 at the indices of the relevant variables
$\{ 1,3,4,5 \}$. For Grover, as there is a unique
input $u$ for which $f(u) = 1$, the ANF formula must include a
subformula matching the binary representation of $u$, and in fact that
subformula is guaranteed to be the shortest one as shown in
Fig.~\ref{fig:Grover}. This solves Grover's unstructured search problem in a deterministic manner. 

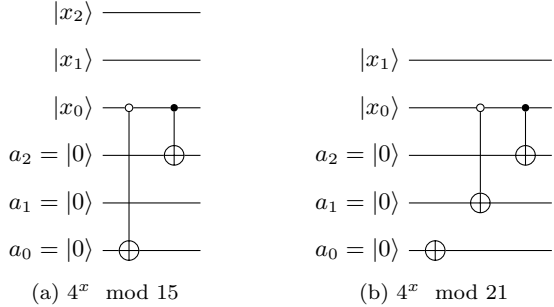
\begin{figure}[H]
\centering
\subfloat[$4^x\mod{15}$\label{fig:shor15}]{%
\begin{tikzpicture}[scale=1.0]
   \begin{yquant*}
   qubit {$\ket{x_2}$} x2;
   qubit {$\ket{x_1}$} x1;
   qubit {$\ket{x_0}$} x0;
   qubit {$a_2 = \ket0$} a2;
   qubit {$a_1 = \ket0$} a1;
   qubit {$a_0 = \ket0$} a0;
   cnot a0 | ~x0; 
   cnot a2 | x0; 
  \end{yquant*}
\end{tikzpicture}
}
\qquad\qquad 
\subfloat[$4^x\mod{21}$\label{fig:shor21}]{%
\begin{tikzpicture}[scale=1.0]
   \begin{yquant*}
   qubit {$\ket{x_1}$} x1;
   qubit {$\ket{x_0}$} x0;
   qubit {$a_2 = \ket0$} a2;
   qubit {$a_1 = \ket0$} a1;
   qubit {$a_0 = \ket0$} a0;
   not a0;
   align -;
   cnot a1 | ~x0; 
   cnot a2 | x0; 
  \end{yquant*}
\end{tikzpicture}
}
\caption{Period finding circuits. (a) The circuit uses qubits
and conventional gates. (b) The circuit  uses
  qutrits. The three gates are from left to right are the $X$,
  $\textrm{SUM}$, and $C(X)$ gates for ternary
  arithmetic~\cite{10.5555/3179473.3179481}. The $X$ gate adds 1
  modulo 3; the controlled version $C(X)$ only increments when the
  control is equal to 2, and the \textrm{SUM} gates maps $\ket{a,b}$
  to $\ket{a,a+b}$.}
\end{figure}

{\it Shor's Algorithm.--}
The circuit in Fig.~\ref{fig:shor15} uses a hand-optimized
implementation of the quantum oracle $U_f$ for the modular exponentiation
function $f(x) = 4^x \mod{15}$. In the graphical representation, an empty circle 
indicates that a negative control bit (active when it is 0) 
and a black circle indicates a positive control bit (active when it is 1). In conventional
forward execution, the state before the QFT block is:
\[
\frac{1}{2\sqrt{2}} (
  (\ket{0} + \ket{2} + \ket{4} + \ket{6}) \ket{1} + 
  (\ket{1} + \ket{3} + \ket{5} + \ket{7}) \ket{4}
  ) .
\]
At this point, the output register is measured to either $\ket{1}$ or
$\ket{4}$. In either case, the input register snaps to a state
of the form $\sum_{r=0}^3 \ket{a+2r}$ whose QFT has peaks at $\ket{0}$
or $\ket{4}$ making them the most likely outcomes of measurements of
the input register. If we measure $\ket{0}$, we repeat the
experiment; otherwise we infer that the period is~2.

\begin{figure*}[t]
\[\begin{array}{l@{\qquad}llll@{\qquad}l}
\textrm{Base} & \multicolumn{4}{c}{\textrm{Equations}} & \textrm{Solution} \\[2ex]
\blue{a=11} & x_0 = 0 &&&& \red{x_0 = 0} \\
\blue{a=4,14} & 1 \oplus x_0 = 1 & x_0 = 0 &&
  & \red{x_0 = 0} \\
\blue{a=7,13} & 1 \oplus x_1 \oplus x_0x_1 = 1 & x_0x_1 = 0 & x_0 \oplus x_1 \oplus x_0x_1 = 0 &  x_0 \oplus x_0x_1 = 0 & \red{x_0=x_1=0} \\
\blue{a=2,8} & 1 \oplus x_0 \oplus x_1 \oplus x_0x_1 = 1 & x_0x_1 = 0 & x_1 \oplus x_0x_1 = 0 & x_0 \oplus x_0x_1 = 0  & \red{x_0=x_1=0} 
\end{array}\]
\caption{Equations generated by retrodictive
  execution of $a^x \mod{15}$ for different values of $a$, starting
  from observed result 1 and unknown
  $x_8x_7x_6x_5x_4x_3x_2x_1x_0$. The solution for the unknown
  variables is given in the last column.}
  \label{fig:shor-eqs}
\end{figure*}

In the retrodictive execution, we can start with the state
$\ket{x_2x_1x_0001}$ since 1 is guaranteed to be a possible output
measurement (corresponding to $f(0)$). The first \cx-gate changes the
state to $\ket{x_2x_1x_0x_001}$ and the second \cx-gate produces
$\ket{x_2x_1x_0x_00x_0}$. At that point, we reconcile the retrodictive
result of the output register $\ket{x_00x_0}$ with the initial
condition $\ket{000}$ to conclude that $x_0=0$. In other words, in
order to observe the output at $001$, the input register must
be initialized to a superposition of the form $\ket{??0}$ where the
least significant bit must be 0 and the other two bits are
unconstrained. Expanding the possibilities, the first register needs
to be in a superposition of the states $\ket{000}, \ket{010},
\ket{100}$ or $\ket{110}$ and we have just inferred using purely
classical but retrodictive reasoning that the period is
2.

This result does not, in fact, require the small optimized circuit of
Fig.~\ref{fig:shor15}. In our implementation, modular exponentiation
circuits are constructed from first principles using adders and
multipliers~\cite{PhysRevA.54.147} and have size $\mathcal{O}(n^3)$ for a circuit with $n$ qubits (Appendix). In the case of $f(x) = 4^x
\mod{15}$, although the unoptimized constructed circuit has 56,538
generalized Toffoli gates (controlled$^{n}$-not gates for all $n$ with both positive and negative controls),
the execution results in just two simple equations: $x_0 = 0$ and $1
\oplus x_0 = 1$. Furthermore, as shown in Fig.~\ref{fig:shor-eqs}, the
shape and size of the equations is largely insensitive to the choice
of 4 as the base of the exponent, leading in all cases to the
immediate conclusion that the period is either 2 or 4. When the
solution is $x_0=0$, the period is 2, and when it is $x_0=x_1=0$, the
period is~4.

The remarkable effectiveness of retrodictive computation of the Shor
instance for factoring 15 is, however, due to a coincidence: a period that is a
power of 2 is trivial to represent in the binary number system
which, after all, is expressly designed for that purpose. That
coincidence repeats itself when factoring products of the (known)
Fermat primes: 3, 5, 17, 257, and 65537, and leads to small
circuits~\cite{shorFermat}. This is confirmed with our implementation
which smoothly deals with unoptimized circuits for factoring such
products. Factoring 3*17=51 using the unoptimized circuit of 177,450
generalized Toffoli gates produces just the 4 equations: $1 \oplus x_1
= 1$, $x_0 = 0$, $x_0 \oplus x_0x_1 = 0$, and $x_1 \oplus x0x1 =
0$. Even for 3*65537=196611 whose circuit has 4,328,778 generalized
Toffoli gates, the execution produces 16 small equations that refer to
just the four variables $x_0$, $x_1$, $x_2$, and $x_3$ constraining
them to be all 0, i.e., asserting that the period is 16.

Since periods that are powers of 2 are rare and special, we turn our
attention to factoring problems with other periods. The simplest such
problem is that of factoring 21 with an underlying function $f(x) =
4^x \mod{21}$ of period 3. The unoptimized circuit constructed from
the first principles has 78,600 generalized Toffoli gates; its
execution generates just three equations. But even in this rather
trivial situation, the equations span 5 pages of text!  (Appendix). A small optimization reducing the number of qubits results
in a circuit of 15,624 generalized Toffoli gates whose execution
produces still quite large, but more reasonable, equations
(Appendix). Despite this blowup, it would however be incorrect to conclude that factoring 21 is
inherently harder than factoring 15. The cause and cure for these unwieldy equations is explained next. 

{\it Complexity Analysis.--} 
Like regular execution, symbolic execution (whether in predictive or retrodictive mode) makes one pass over the circuit, touching 
each gate once. Thus the cost of symbolic execution is entirely dominated by the amount of work
done when processing an individual gate. This cost is proportional to the size of the formulae representing the inputs to the gate and as the example of Shor 21 shows, this size can be exponentially large. On one hand, this
exponential blowup can be viewed as a good sign as exponentially large intermediate states are a necessary condition for any quantum
algorithm that offers an exponential speed-up over classical
computation~\cite{10.2307/3560059}. We can still ask, however, if perhaps the exponential size is due to our choice of ANF and whether another representation could be more efficient. Indeed, it is clear that the issue is simply that the
binary number system is well-tuned to expressing patterns over powers
of 2 (yielding small formulae in those cases) but a very poor match for expressing patterns over powers of other prime numbers 
(yielding large formulae in those cases). This suggests that by using qutrits, the circuit and
equations for factoring 21 would become trivial. And indeed, the circuit in
Fig.~\ref{fig:shor21} for the modular exponentiation routine needed for factoring 21 consists of just three gates; its retrodictive
execution produces two equations: $x_0=0$ and $x_0 \neq 2$, setting
$x_0=0$ and leaving $x_1$ unconstrained. The matching values in the
qutrit system are 00, 10, 20 or in decimal 0, 3, 6 identifying
the period to be~3.

\begin{figure}[!]
\includegraphics[scale=0.35]{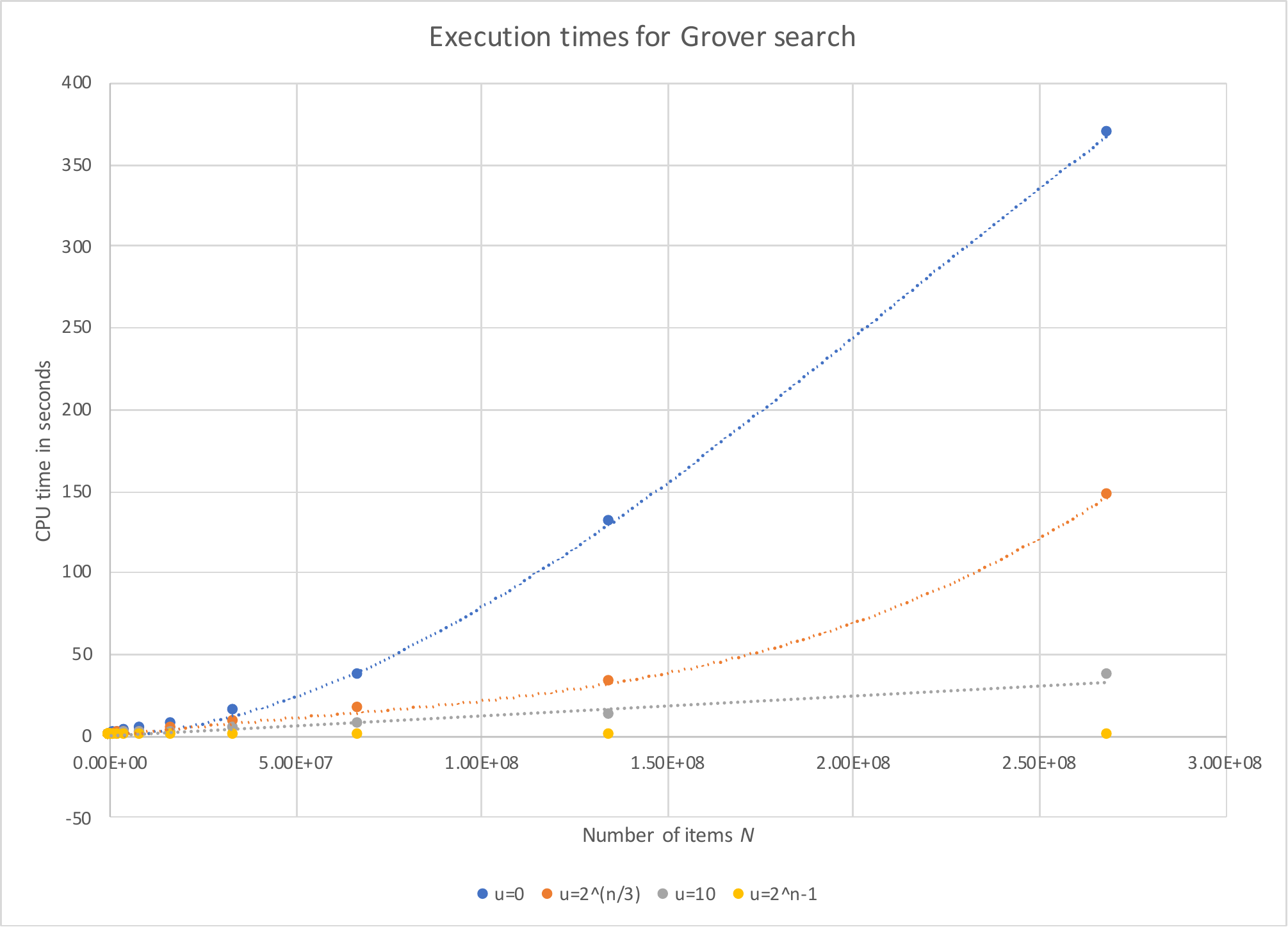}
\caption{\label{fig:grover}Execution times for Grover's search for various numbers of elements $N$ and for various choices of the marked element $u$}
\end{figure}

Even if we uniformly maintain the ANF representation, retrodictive and symbolic executions are surprisingly efficient for some instances of the quantum algorithms we discussed. Most significantly, a Grover function $f : \finset{2^\mathit{n}} \rightarrow \finset{2}$ such that $f(u) = 1$ for a unique marked input $u$ induces many possible implementations of $U_f$ that are all equivalent to a circuit with a \emph{single} generalized Toffoli gate that negates the output qubit when the controls match~$u$. Together with the fact that we only need the shortest subformula to identify~$u$, instances of Grover search are all solved in a single pass (in our white-box complexity model), as opposed to $\mathcal{O}(\sqrt{2^n})$ queries. The performance depends on the number of zeros in the binary representation of the marked element.  Fig.~\ref{fig:grover} shows the execution times for various choices of the marked element $u$ as a function of the number of elements $N$. The worst case occurs when $u=0$ and is proportional to $N$. In the best case, the marked element is $2^n-1$ (all ones in binary) and the symbolic retrodictive execution time is negligible far beyond the scale of the chart, remaining less then 30 milliseconds on a conventional laptop even as $n$ reaches 1000, i.e., searching among $N=2^{1000}$ items. 

In summary, the reasons for the efficacy of symbolic retrodictive execution can be recapped as: (i) retrodictive execution, classical or quantum, is inherently optimized to consider a single output of interest alleviating the need to consider inputs leading to other values of the output register when analyzing many-to-one functions, (ii) for many quantum algorithms, the symbolic formulae need not be explicitly solved and in fact are not even needed in their full representation as the solution can in some cases be gleaned from parts of the formulae since it represents a relational property of the function's output, (iii) for many quantum algorithms there is neither a need for relative phases (other than a fixed phase zero) nor for measurements, making the symbolic representation adequate for representing the emergent entanglement patterns, and (iv) with judicious adaptive choice of the representation of formulae, the hidden entanglement patterns can be made to disappear as illustrated by the use of qutrits to represent formulae with period 3. This idea of adapting the representation of the computation to simplify
the circuit and equations is inspired by the fact that entanglement is
relative to a particular tensor product decomposition or, more generically, 
to a distinguished subspace of observables~\cite{GE2004} (\hyperref[sec:Methods]{Appendix}). It is therefore
tempting to conjecture that quantum advantage relies on the dynamic reconfiguration
of the computational subspaces. This can certainly be efficiently achieved via the QFT, 
a suspected source of quantum advantage~\cite{haduniv}, 
but it also raises the question of whether, and to what extent, such dynamic
adaptability may be automatically realized classically? Clearly it is all about representation of the flow 
information, one of the best protected secrets of Nature. 

%%%%%%%%%%%%%%%%%%%%%%%%%%%%%%%%%%%%%%%%%%%%%%%%%%%%%%%%%%%%%%%%%%%%%%%%%%%%%%%%%%%%%%%%%%
\vspace*{-0.5cm}
\section*{Appendix}\label{sec:Methods}

\paragraph*{Quantum Algorithms.}  
Standard quantum gates include Hadamard, $H=\frac{1}{\sqrt{2}}\begin{pmatrix}1& 1 \\ 1& -1 \end{pmatrix}$, 
and controlled-not (\cx) gate, \cx $=\begin{pmatrix}1& 0 \\ 0 & 1 \end{pmatrix}\oplus 
\begin{pmatrix}0& 1 \\ 1& 0 \end{pmatrix}$, or \cx $= 
\begin{pmatrix}0& 1 \\ 1& 0 \end{pmatrix}\oplus \begin{pmatrix}1& 0 \\ 0 & 1 \end{pmatrix}$, depending on 
the control bit being positive (1) or negative (0), respectively. 
In the \textbf{Deutsch} problem, we are given a function $\finset{2} \rightarrow \finset{2}$;
the goal is to determine if the function is constant or balanced.
The \textbf{Deutsch-Jozsa} problem generalizes this question to functions of type
$\finset{2^\mathit{n}} \rightarrow \finset{2}$.
In the \textbf{Bernstein-Vazirani} problem, we are given a function $\finset{2^\mathit{n}} \rightarrow \finset{2}$ 
that hides a secret number $s \in \finset{2^\mathit{n}}$. We are promised that the function is defined using the 
binary representations $\sum_i^{n-1} x_i$ and $\sum_i^{n-1} s_i$ of $x$ and $s$ respectively as 
$f(x) = \sum_{i=0}^{n-1} s_ix_i \mod{2}$. The goal is to determine the secret number $s$.
In the \textbf{Simon} problem, we are given a 2-1 function $f : \finset{2^\mathit{n}} \rightarrow \finset{2^\mathit{n}}$
with the property that there exists an $a$ such that $f(x) = f(x \oplus a)$ for all $x$; the goal is
to determine $a$. In \textbf{Grover}'s algorithm, we are given a function $f : \finset{2^\mathit{n}} \rightarrow \finset{2}$
with the property that there exists only one input $u$ such that $f(u) = 1$; the goal is to find $u$.
In \textbf{Shor}'s algorithm, we are given a number $N$ to factor and a
function $f : \finset{Q} \rightarrow \finset{Q}$ where $Q = \lceil \log_2 N^2 \rceil$ and where
$f(x) = a^x \mod{N}$. For appropriate values of~$a$, this function has the property that there exists an $r$ such 
that $f(x) = f(x+r)$ for all $x$; the goal is to determine $r$. 

\paragraph*{Symbolic Execution of Classical Programs.}
A well-established technique to simultaneously explore multiple paths
that a classical program could take under different inputs is
\emph{symbolic
  execution}~\cite{10.1145/390016.808445,10.1145/360248.360252,howden,10.1145/800191.805647,10.1145/3182657}. In
this execution scheme, concrete values are replaced by symbols which are
initially unconstrained. As the execution proceeds, the symbols
interact with program constructs and this typically introduces
constraints on the possible values that the symbols represent. In this way, one 
builds {\it classical} correlations among symbolic variables. At the
end of the execution, these constraints can be solved to infer
properties of the program under consideration. 

\paragraph*{Algebraic Normal Form (ANF).}
The semantics of a generalized Toffoli gate with $n$ positive control qubits:
$a_{n-1},\cdots,a_0$ and one target qubit $b$ is $b \oplus \bigwedge_i
a_i$, the exclusive-or of the target $b$ with the conjunction of all
the control qubits. This form is precisely the definition of the algebraic
normal form of boolean expressions. 

When all the variables are bits, the execution of one 
gate takes time proportional to the number of control wires. But when the 
variables range over formulae, the execution of one gate additionally depends
on the size of formulae. As each conjunction has the potential of doubling the 
size of the formula, this can quickly get prohibitively expensive. We note that circuits that only
use \x\ and \cx-gates never generate any conjunctions and hence lead
to formulae that are efficiently solvable
classically~\cite{10.5555/35517,TOKAREVA20151}. This situation is reminiscent of the Gottesman-Knill theorem \cite{GKThm}.

There are two properties of ANF that make it an excellent representation for symbolic formulae. First, being a normal form, the ANF of a circuit is unique which
explains why wildly different circuits for the same algorithm yield the same
equations. Additionally, a general ANF formula of the form $ X_1 \oplus
X_2 \oplus X_3 \oplus \ldots = 0$ where each $X_i$ is a conjunction of
some boolean variables can express both constructive and destructive interference.
All the variables in each $X$ must all be true to enable that
$X=1$, i.e., they constructively interfere. But since the entire formula must equal to 0, every $X_i = 1$ must
be offset by another $X_j = 1$, thus exhibiting destructive interference
among $X_i$ and $X_j$. Generally speaking, arbitrary interference
patterns can be encoded in the formulae at the cost of making the size
of the formulae exponential in the number of variables.

\begin{figure}[t]
\begin{center}
\begin{tikzpicture}[scale=0.8]
   \begin{yquant*}[register/minimum height=1.3cm]
   qubit {$\ket0$} x;
   qubit {$\ket0$} y;
   box {$H$} x;
   cnot y | x;
  \end{yquant*}
\end{tikzpicture}
\qquad\qquad
\begin{tikzpicture}[scale=0.8]
   \begin{yquant*}[register/minimum height=1.3cm]
   qubit {$\ket0$} x;
   qubit {$\ket0$} y;
   qubit {$\ket0$} z;
   box {$H$} x;
   cnot y | x;
   cnot z | y;
  \end{yquant*}
\end{tikzpicture}
\end{center}
\caption{\label{fig:bell2}Bell and GHZ States}
\end{figure}
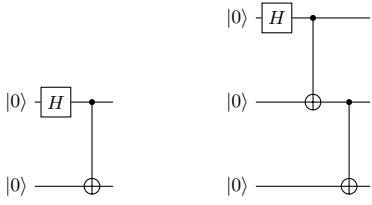
\paragraph*{Entanglement.}
A symbolic variable represents a boolean value that can be 0 or 1;
this is similar to a qubit in a superposition $(1/\sqrt{2}) (\ket{0}
\pm \ket{1})$. Thus, it appears that $H\ket{0}$ could be represented
by a symbol~$x$ to denote the uncertainty. Surprisingly, this idea
scales to even represent maximally entangled
states. Fig.~\ref{fig:bell2} (left) shows a circuit to generate the Bell
state $(1/\sqrt{2}) (\ket{00} + \ket{11})$. By using the symbol $x$
for $H\ket{0}$, the input to the \cx-gate is $\ket{x0}$ which
evolves to $\ket{xx}$. By sharing the same symbol in two positions,
the symbolic state accurately represents the entangled Bell
state. Similarly, for the circuit in Fig.~\ref{fig:bell2} (right), the
state after the Hadamard gate is $\ket{x00}$ which evolves to
$\ket{xx0}$ and then to $\ket{xxx}$ again accurately capturing the
entanglement correlations.

Given a maximally entangled state defined with respect to a particular
tensor product decomposition, the same state may become unentangled in
a different tensor product decomposition. Given the (unnormalized) state:
\[
 \ket{\Psi} =\ket{0}+\ket{3}+\ket{6}+\ket{9}+\ket{12}+\ket{15},
\]
one can find a 4-qubit representation (${\cal H}=\bigotimes_{i=1}^4
\mathbb{C}^2$)
\[
 \ket{\Psi} = \ket{0000}+\ket{0011}+\ket{0110} 
  + \ket{1001}+\ket{1100}+\ket{1111},
\]
where we used the following map $\ket{m}=\sum_{i=0}^3 x_i 2^i$, with
$m \in \mathbb{Z}$ and $x_i=0,1$.  One can use the purity~\cite{GE2004}
\[
 P_{\ket{\Psi} }=\frac{1}{4}\sum_{i=1}^4\sum_{\mu=x,y,z}\langle \Psi |
 \sigma^\mu_i \ket{\Psi}^2 ,
\]
where $\sigma^\mu_i$ are Pauli matrices, 
and confirm that the state $\ket{\Psi}$ is maximally entangled, i.e.,
has $P_{\ket{\Psi}} =0$. In contrast, in a qutrit basis (${\cal
  H}=\bigotimes_{i=1}^4 \mathbb{C}^3$), given the map
$\ket{m}=\sum_{i=0}^3 x_i 3^i$, with $x_i=0,1,2$, the state
\begin{eqnarray}
 \ket{\Psi} &=&\ket{0000}+\ket{0010}+\ket{0020} 
  + \ket{0100}+ \nonumber \\
  && \quad\ket{0110}+\ket{0120} \nonumber \\
 &=&\ket{0}\otimes(\ket{0}+\ket{1})\otimes(\ket{0}+\ket{1}+\ket{2})\otimes\ket{0}, \nonumber
\end{eqnarray}
is a product (unentangled) state. 

\paragraph*{Partial Evaluation.}
Below is a Haskell~\cite{haskell} program that computes $a^n$ by repeated squaring:
\begin{minted}{haskell}
power :: Int -> Int -> Int
power a n
  | n == 0     = 1
  | n == 1     = a
  | even n     = let r = power a (n `div` 2) 
                 in r * r 
  | otherwise  = a * power a (n-1)
\end{minted}
When both inputs are known, e.g., \h{a = 3} and \h{n = 5}  the
program evaluates as follows:
\begin{minted}{haskell}
   power 3 5
=  3 * power 3 4
=  3 * (let r1 = power 3 2 in r1 * r1)
=  3 * (let r1 = 
         (let r2 = power 3 1 in r2 * r2) 
          in r1 * r1)
=  3 * (let r1 = 
         (let r2 = 3 in r2 * r2) in r1 * r1)
=  3 * (let r1 = 9 in r1 * r1)
=  243
\end{minted}

Partial evaluation is used when we only have partial information about
the inputs. Say we only know $n=5$. A partial evaluator then attempts
to evaluate \h{power} with symbolic input \h{a} and actual input
\h{n=5}  This evaluation proceeds as follows:
\begin{minted}{haskell}
   power a 5 
=  a * power a 4 
=  a * (let r1 = power a 2 in r1 * r1)
=  a * (let r1 = 
         (let r2 = power a 1 in r2 * r2) 
          in r1 * r1)
=  a * (let r1 = 
         (let r2 = a in r2 * r2) 
          in r1 * r1)
=  a * (let r1 = a * a in r1 * r1)
=  let r1 = a * a in a * r1 * r1
\end{minted}
All of this evaluation, simplification, and specialization happens
without knowledge of \h{a}. Just knowing~\h{n} was enough to
produce a residual program that is much simpler. This is the underlying technology for 
our symbolic evaluator.

\begin{figure}[htb]
\begin{center}
\begin{tikzpicture}[scale=0.6]
\draw (0,0) ellipse (1cm and 2cm);
\draw (5,0) circle (1cm and 2cm);
\fill[blue!20!white] (0,0.6) ellipse (0.5cm and 1.2cm);
\node (a) at (0,1.45) {2};
\node (b) at (0,0.95) {6};
\node (c) at (0,0.45) {10};
\node (d) at (0,-0.05) {14};
\node (t) at (5,0.5) {4};
\node at (0,-1) {$\vdots$};
\node at (5,-1) {$\vdots$};
\draw[->] (a) to node [above] {$f$} (t);
\draw[->] (b) -- (t);
\draw[->] (c) -- (t);
\draw[->] (d) -- (t);
\end{tikzpicture}
\end{center}
\caption{\label{fig:preimage}Pre-image of 4 under $f(x) = 7^x \mod 15$.}
\end{figure}
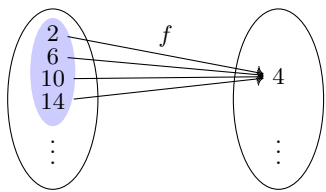
\paragraph*{Complexity Analysis.}
Given finite sets $A$ and $B$, a function $f : A \rightarrow B$ and an
element $y \in B$, we define $\preim{f}{y}$, the pre-image of~$y$
under~$f$, as the set $\{ x \in A ~|~ f(x) = y \}$. For example, let
$A = B = \finset{2^\mathrm{4}}$ and let $f(x) = 7^x \mod 15$, then the
collection of values that $f$ maps to~4, $\preim{f}{4}$, is the set
$\{ 2, 6, 10, 14 \}$ as shown in Fig.~\ref{fig:preimage}. Symbolic
retrodictive execution can be seen as a method to generate boolean
formulae that describe the pre-image of the function $f$ under
study. For the example in Fig.~\ref{fig:preimage}, retrodictive
execution might generate the formulae $x_1=1$ and $x_0=0$. The
(trivial in this case) solution for the formulae is indeed the set $\{
2, 6, 10, 14 \}$. The critical points to note, however, are that: (i)
solving the equations describing the pre-image is in general an
intractable (even for quantum computers) $\mathit{NP}$-complete
problem, and (ii) solving the equations is not needed for typical
quantum algorithms. \emph{Only some global or relational properties of the pre-image
  are needed!} Indeed, we have already seen that for solving the
Deutsch-Jozsa problem, the only thing needed was whether the formula
contains some variables. For the Bernstein-Vazirani problem, the only
thing needed was the indices of the variables occurring in the
formula. For Grover's algorithm, we only need to extract the singleton
element in the pre-image and for Shor's algorithm we ``only'' need to
extract the periodicity of the elements in the pre-image.

To appreciate the difficulty of computing pre-images in general, note
that finding the pre-image of a function subsumes several challenging
computational problems such as pre-image attacks on hash
functions~\cite{10.1007/978-3-540-25937-4_24}, predicting
environmental conditions that allow certain reactions to take place in
computational biology~\cite{Klotz2013,akutsu2009analyses}, and finding
the pre-image of feature vectors in the space induced by a kernel in
neural networks~\cite{1353287}. More to the point, the boolean
satisfiability problem SAT is expressible as a boolean function over
the input variables and solving a SAT problem is asking for the
pre-image of \textsf{true}. Indeed, based on the conjectured existence
of one-way functions which itself implies $\mathit{P} \neq
\mathit{NP}$, all these pre-images calculations are believed to be
computationally intractable in their most general setting.

\begin{figure}[b]
    \centering
    \includegraphics[scale=0.6]{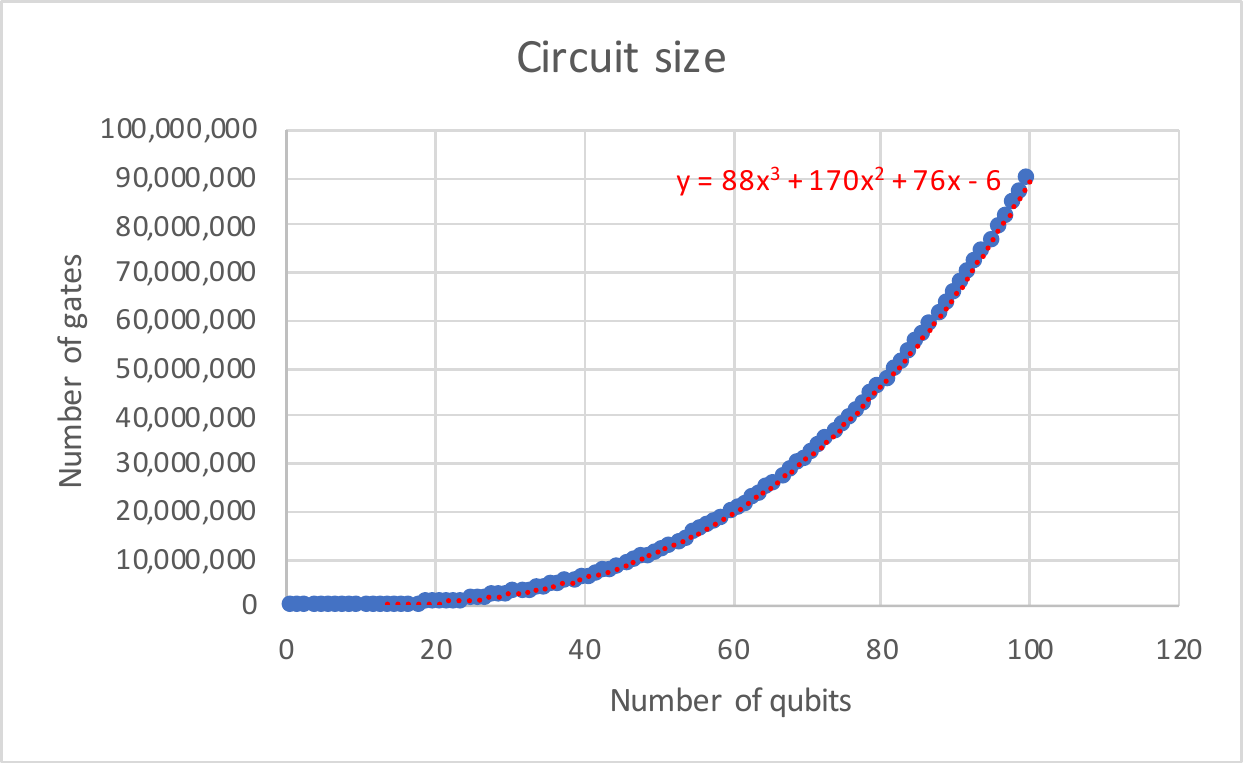}
    \caption{Number of gates in modular exponentiation circuits as a function of the number of qubits. The blue dots are the actual counts. The red line is the fitted equation (displayed in red)}
    \label{fig:gates}
\end{figure}
\paragraph*{Software.}
The entire suite of programs including synthesis of reversible
circuits, standard evaluation, retrodictive evaluation under various
modes, testing, debugging, and alternative representations of ANF
formulae is about 1,500 lines of Haskell. Fig.~\ref{fig:gates} shows the number of gates in the unoptimized modular exponentiation circuits. The heart of the
implementation is this simple function:

\begin{minted}{haskell}
peG :: Value v => GToffoli s v -> ST s ()
peG (GToffoli bs cs t) = do
  controls <- mapM readSTRef cs
  tv <- readSTRef t
  let funs = 
    map (\b -> if b then id else snot) bs
  let r = 
    sxor tv 
      (foldr sand one 
        (zipWith ($) funs controls))
  writeSTRef t r
\end{minted}
The function performs symbolic evaluation of one generalized Toffoli
gate, reading the current ANF formulae for each control and producing
an appropriate ANF formula for the target.

\paragraph*{Data Availability.}
All execution results will be made available and can be replicated by
executing the associated software.

\paragraph*{Code Availability.}
The computer programs used to generate the circuits and symbolically
execute the quantum algorithms retrodictively will be made publicly
available.

\paragraph*{Author Contributions.}
The idea of symbolic evaluation is due to A.S. The connection to
retrodictive quantum mechanics is due to G.O. The connection to
partial evaluation is due to J.C. Both A.S. and J.C. contributed to
the software code to run the experiments. Both A.S. and
G.O. contributed to the analysis of the quantum algorithms and their
de-quantization. All authors contributed to the writing of the
document.

\paragraph*{Competing Interests.}
No competing interests.

\paragraph*{Materials \& Correspondence.}
The corresponding author is Gerardo Ortiz. 

\onecolumngrid
\bigskip

\paragraph*{Equations for Shor 21.}
\label{par:shor21}

The equations generated by retrodictive execution of the optimized
circuit for $4^x \mod{21}$ starting from observed result 1 and unknown
$x$ are:

\bigskip

$1 \oplus x_0 \oplus x_1 \oplus x_2 \oplus x_0x_2 \oplus x_0x_1x_2
\oplus x_3 \oplus x_1x_3 \oplus x_0x_1x_3 \oplus x_0x_2x_3 \oplus
x_1x_2x_3 \oplus x_4 \oplus x_0x_4 \oplus x_0x_1x_4 \oplus x_2x_4
\oplus x_1x_2x_4 \oplus x_0x_1x_2x_4 \oplus x_0x_3x_4 \oplus x_1x_3x_4
\oplus x_2x_3x_4 \oplus x_0x_2x_3x_4 \oplus x_0x_1x_2x_3x_4 \oplus x_5
\oplus x_1x_5 \oplus x_0x_1x_5 \oplus x_0x_2x_5 \oplus x_1x_2x_5
\oplus x_3x_5 \oplus x_0x_3x_5 \oplus x_0x_1x_3x_5 \oplus x_2x_3x_5
\oplus x_1x_2x_3x_5 \oplus x_0x_1x_2x_3x_5 \oplus x_0x_4x_5 \oplus
x_1x_4x_5 \oplus x_2x_4x_5 \oplus x_0x_2x_4x_5 \oplus x_0x_1x_2x_4x_5
\oplus x_3x_4x_5 \oplus x_1x_3x_4x_5 \oplus x_0x_1x_3x_4x_5 \oplus
x_0x_2x_3x_4x_5 \oplus x_1x_2x_3x_4x_5 = 1$

\bigskip

$x_1 \oplus x_0x_1 \oplus x_0x_2 \oplus x_1x_2 \oplus x_3 \oplus
x_0x_3 \oplus x_0x_1x_3 \oplus x_2x_3 \oplus x_1x_2x_3 \oplus
x_0x_1x_2x_3 \oplus x_0x_4 \oplus x_1x_4 \oplus x_2x_4 \oplus
x_0x_2x_4 \oplus x_0x_1x_2x_4 \oplus x_3x_4 \oplus x_1x_3x_4 \oplus
x_0x_1x_3x_4 \oplus x_0x_2x_3x_4 \oplus x_1x_2x_3x_4 \oplus x_5 \oplus
x_0x_5 \oplus x_0x_1x_5 \oplus x_2x_5 \oplus x_1x_2x_5 \oplus
x_0x_1x_2x_5 \oplus x_0x_3x_5 \oplus x_1x_3x_5 \oplus x_2x_3x_5 \oplus
x_0x_2x_3x_5 \oplus x_0x_1x_2x_3x_5 \oplus x_4x_5 \oplus x_1x_4x_5
\oplus x_0x_1x_4x_5 \oplus x_0x_2x_4x_5 \oplus x_1x_2x_4x_5 \oplus
x_3x_4x_5 \oplus x_0x_3x_4x_5 \oplus x_0x_1x_3x_4x_5 \oplus
x_2x_3x_4x_5 \oplus x_1x_2x_3x_4x_5 \oplus x_0x_1x_2x_3x_4x_5 = 0$

\bigskip

$x_0 \oplus x_0x_1 \oplus x_2 \oplus x_1x_2 \oplus x_0x_1x_2 \oplus
x_0x_3 \oplus x_1x_3 \oplus x_2x_3 \oplus x_0x_2x_3 \oplus
x_0x_1x_2x_3 \oplus x_4 \oplus x_1x_4 \oplus x_0x_1x_4 \oplus
x_0x_2x_4 \oplus x_1x_2x_4 \oplus x_3x_4 \oplus x_0x_3x_4 \oplus
x_0x_1x_3x_4 \oplus x_2x_3x_4 \oplus x_1x_2x_3x_4 \oplus
x_0x_1x_2x_3x_4 \oplus x_0x_5 \oplus x_1x_5 \oplus x_2x_5 \oplus
x_0x_2x_5 \oplus x_0x_1x_2x_5 \oplus x_3x_5 \oplus x_1x_3x_5 \oplus
x_0x_1x_3x_5 \oplus x_0x_2x_3x_5 \oplus x_1x_2x_3x_5 \oplus x_4x_5
\oplus x_0x_4x_5 \oplus x_0x_1x_4x_5 \oplus x_2x_4x_5 \oplus
x_1x_2x_4x_5 \oplus x_0x_1x_2x_4x_5 \oplus x_0x_3x_4x_5 \oplus
x_1x_3x_4x_5 \oplus x_2x_3x_4x_5 \oplus x_0x_2x_3x_4x_5 \oplus
x_0x_1x_2x_3x_4x_5 = 0$

\bigskip

The equations generated by retrodictive execution of the unoptimized
$4^x \mod{21}$ starting from observed result 1 and unknown $x$. The
circuit consists of 36,400 \cx-gates, 38,200 \ccx-gates, and 4,000
\cccx-gates. There are only three equations but each equation is
exponentially large:

\bigskip

$1 \oplus x0 \oplus x_1 \oplus x_2 \oplus x_0x_2 \oplus x_0x_1x_2
\oplus x_3 \oplus x_1x_3 \oplus x_0x_1x_3 \oplus x_0x_2x_3 \oplus
x_1x_2x_3 \oplus x_4 \oplus x_0x_4 \oplus x_0x_1x_4 \oplus x_2x_4
\oplus x_1x_2x_4 \oplus x_0x_1x_2x_4 \oplus x_0x_3x_4 \oplus x_1x_3x_4
\oplus x_2x_3x_4 \oplus x_0x_2x_3x_4 \oplus x_0x_1x_2x_3x_4 \oplus x_5
\oplus x_1x_5 \oplus x_0x_1x_5 \oplus x_0x_2x_5 \oplus x_1x_2x_5
\oplus x_3x_5 \oplus x_0x_3x_5 \oplus x_0x_1x_3x_5 \oplus x_2x_3x_5
\oplus x_1x_2x_3x_5 \oplus x_0x_1x_2x_3x_5 \oplus x_0x_4x_5 \oplus
x_1x_4x_5 \oplus x_2x_4x_5 \oplus x_0x_2x_4x_5 \oplus x_0x_1x_2x_4x_5
\oplus x_3x_4x_5 \oplus x_1x_3x_4x_5 \oplus x_0x_1x_3x_4x_5 \oplus
x_0x_2x_3x_4x_5 \oplus x_1x_2x_3x_4x_5 \oplus x_6 \oplus x_0x_6 \oplus
x_0x_1x_6 \oplus x_2x_6 \oplus x_1x_2x_6 \oplus x_0x_1x_2x_6 \oplus
x_0x_3x_6 \oplus x_1x_3x_6 \oplus x_2x_3x_6 \oplus x_0x_2x_3x_6 \oplus
x_0x_1x_2x_3x_6 \oplus x_4x_6 \oplus x_1x_4x_6 \oplus x_0x_1x_4x_6
\oplus x_0x_2x_4x_6 \oplus x_1x_2x_4x_6 \oplus x_3x_4x_6 \oplus
x_0x_3x_4x_6 \oplus x_0x_1x_3x_4x_6 \oplus x_2x_3x_4x_6 \oplus
x_1x_2x_3x_4x_6 \oplus x_0x_1x_2x_3x_4x_6 \oplus x_0x_5x_6 \oplus
x_1x_5x_6 \oplus x_2x_5x_6 \oplus x_0x_2x_5x_6 \oplus x_0x_1x_2x_5x_6
\oplus x_3x_5x_6 \oplus x_1x_3x_5x_6 \oplus x_0x_1x_3x_5x_6 \oplus
x_0x_2x_3x_5x_6 \oplus x_1x_2x_3x_5x_6 \oplus x_4x_5x_6 \oplus
x_0x_4x_5x_6 \oplus x_0x_1x_4x_5x_6 \oplus x_2x_4x_5x_6 \oplus
x_1x_2x_4x_5x_6 \oplus x_0x_1x_2x_4x_5x_6 \oplus x_0x_3x_4x_5x_6
\oplus x_1x_3x_4x_5x_6 \oplus x_2x_3x_4x_5x_6 \oplus
x_0x_2x_3x_4x_5x_6 \oplus x_0x_1x_2x_3x_4x_5x_6 \oplus x_7 \oplus
x_1x_7 \oplus x_0x_1x_7 \oplus x_0x_2x_7 \oplus x_1x_2x_7 \oplus
x_3x_7 \oplus x_0x_3x_7 \oplus x_0x_1x_3x_7 \oplus x_2x_3x_7 \oplus
x_1x_2x_3x_7 \oplus x_0x_1x_2x_3x_7 \oplus x_0x_4x_7 \oplus x_1x_4x_7
\oplus x_2x_4x_7 \oplus x_0x_2x_4x_7 \oplus x_0x_1x_2x_4x_7 \oplus
x_3x_4x_7 \oplus x_1x_3x_4x_7 \oplus x_0x_1x_3x_4x_7 \oplus
x_0x_2x_3x_4x_7 \oplus x_1x_2x_3x_4x_7 \oplus x_5x_7 \oplus x_0x_5x_7
\oplus x_0x_1x_5x_7 \oplus x_2x_5x_7 \oplus x_1x_2x_5x_7 \oplus
x_0x_1x_2x_5x_7 \oplus x_0x_3x_5x_7 \oplus x_1x_3x_5x_7 \oplus
x_2x_3x_5x_7 \oplus x_0x_2x_3x_5x_7 \oplus x_0x_1x_2x_3x_5x_7 \oplus
x_4x_5x_7 \oplus x_1x_4x_5x_7 \oplus x_0x_1x_4x_5x_7 \oplus
x_0x_2x_4x_5x_7 \oplus x_1x_2x_4x_5x_7 \oplus x_3x_4x_5x_7 \oplus
x_0x_3x_4x_5x_7 \oplus x_0x_1x_3x_4x_5x_7 \oplus x_2x_3x_4x_5x_7
\oplus x_1x_2x_3x_4x_5x_7 \oplus x_0x_1x_2x_3x_4x_5x_7 \oplus
x_0x_6x_7 \oplus x_1x_6x_7 \oplus x_2x_6x_7 \oplus x_0x_2x_6x_7 \oplus
x_0x_1x_2x_6x_7 \oplus x_3x_6x_7 \oplus x_1x_3x_6x_7 \oplus
x_0x_1x_3x_6x_7 \oplus x_0x_2x_3x_6x_7 \oplus x_1x_2x_3x_6x_7 \oplus
x_4x_6x_7 \oplus x_0x_4x_6x_7 \oplus x_0x_1x_4x_6x_7 \oplus
x_2x_4x_6x_7 \oplus x_1x_2x_4x_6x_7 \oplus x_0x_1x_2x_4x_6x_7 \oplus
x_0x_3x_4x_6x_7 \oplus x_1x_3x_4x_6x_7 \oplus x_2x_3x_4x_6x_7 \oplus
x_0x_2x_3x_4x_6x_7 \oplus x_0x_1x_2x_3x_4x_6x_7 \oplus x_5x_6x_7
\oplus x_1x_5x_6x_7 \oplus x_0x_1x_5x_6x_7 \oplus x_0x_2x_5x_6x_7
\oplus x_1x_2x_5x_6x_7 \oplus x_3x_5x_6x_7 \oplus x_0x_3x_5x_6x_7
\oplus x_0x_1x_3x_5x_6x_7 \oplus x_2x_3x_5x_6x_7 \oplus
x_1x_2x_3x_5x_6x_7 \oplus x_0x_1x_2x_3x_5x_6x_7 \oplus x_0x_4x_5x_6x_7
\oplus x_1x_4x_5x_6x_7 \oplus x_2x_4x_5x_6x_7 \oplus
x_0x_2x_4x_5x_6x_7 \oplus x_0x_1x_2x_4x_5x_6x_7 \oplus x_3x_4x_5x_6x_7
\oplus x_1x_3x_4x_5x_6x_7 \oplus x_0x_1x_3x_4x_5x_6x_7 \oplus
x_0x_2x_3x_4x_5x_6x_7 \oplus x_1x_2x_3x_4x_5x_6x_7 \oplus x_8 \oplus
x_0x_8 \oplus x_0x_1x_8 \oplus x_2x_8 \oplus x_1x_2x_8 \oplus
x_0x_1x_2x_8 \oplus x_0x_3x_8 \oplus x_1x_3x_8 \oplus x_2x_3x_8 \oplus
x_0x_2x_3x_8 \oplus x_0x_1x_2x_3x_8 \oplus x_4x_8 \oplus x_1x_4x_8
\oplus x_0x_1x_4x_8 \oplus x_0x_2x_4x_8 \oplus x_1x_2x_4x_8 \oplus
x_3x_4x_8 \oplus x_0x_3x_4x_8 \oplus x_0x_1x_3x_4x_8 \oplus
x_2x_3x_4x_8 \oplus x_1x_2x_3x_4x_8 \oplus x_0x_1x_2x_3x_4x_8 \oplus
x_0x_5x_8 \oplus x_1x_5x_8 \oplus x_2x_5x_8 \oplus x_0x_2x_5x_8 \oplus
x_0x_1x_2x_5x_8 \oplus x_3x_5x_8 \oplus x_1x_3x_5x_8 \oplus
x_0x_1x_3x_5x_8 \oplus x_0x_2x_3x_5x_8 \oplus x_1x_2x_3x_5x_8 \oplus
x_4x_5x_8 \oplus x_0x_4x_5x_8 \oplus x_0x_1x_4x_5x_8 \oplus
x_2x_4x_5x_8 \oplus x_1x_2x_4x_5x_8 \oplus x_0x_1x_2x_4x_5x_8 \oplus
x_0x_3x_4x_5x_8 \oplus x_1x_3x_4x_5x_8 \oplus x_2x_3x_4x_5x_8 \oplus
x_0x_2x_3x_4x_5x_8 \oplus x_0x_1x_2x_3x_4x_5x_8 \oplus x_6x_8 \oplus
x_1x_6x_8 \oplus x_0x_1x_6x_8 \oplus x_0x_2x_6x_8 \oplus x_1x_2x_6x_8
\oplus x_3x_6x_8 \oplus x_0x_3x_6x_8 \oplus x_0x_1x_3x_6x_8 \oplus
x_2x_3x_6x_8 \oplus x_1x_2x_3x_6x_8 \oplus x_0x_1x_2x_3x_6x_8 \oplus
x_0x_4x_6x_8 \oplus x_1x_4x_6x_8 \oplus x_2x_4x_6x_8 \oplus
x_0x_2x_4x_6x_8 \oplus x_0x_1x_2x_4x_6x_8 \oplus x_3x_4x_6x_8 \oplus
x_1x_3x_4x_6x_8 \oplus x_0x_1x_3x_4x_6x_8 \oplus x_0x_2x_3x_4x_6x_8
\oplus x_1x_2x_3x_4x_6x_8 \oplus x_5x_6x_8 \oplus x_0x_5x_6x_8 \oplus
x_0x_1x_5x_6x_8 \oplus x_2x_5x_6x_8 \oplus x_1x_2x_5x_6x_8 \oplus
x_0x_1x_2x_5x_6x_8 \oplus x_0x_3x_5x_6x_8 \oplus x_1x_3x_5x_6x_8
\oplus x_2x_3x_5x_6x_8 \oplus x_0x_2x_3x_5x_6x_8 \oplus
x_0x_1x_2x_3x_5x_6x_8 \oplus x_4x_5x_6x_8 \oplus x_1x_4x_5x_6x_8
\oplus x_0x_1x_4x_5x_6x_8 \oplus x_0x_2x_4x_5x_6x_8 \oplus
x_1x_2x_4x_5x_6x_8 \oplus x_3x_4x_5x_6x_8 \oplus x_0x_3x_4x_5x_6x_8
\oplus x_0x_1x_3x_4x_5x_6x_8 \oplus x_2x_3x_4x_5x_6x_8 \oplus
x_1x_2x_3x_4x_5x_6x_8 \oplus x_0x_1x_2x_3x_4x_5x_6x_8 \oplus x_0x_7x_8
\oplus x_1x_7x_8 \oplus x_2x_7x_8 \oplus x_0x_2x_7x_8 \oplus
x_0x_1x_2x_7x_8 \oplus x_3x_7x_8 \oplus x_1x_3x_7x_8 \oplus
x_0x_1x_3x_7x_8 \oplus x_0x_2x_3x_7x_8 \oplus x_1x_2x_3x_7x_8 \oplus
x_4x_7x_8 \oplus x_0x_4x_7x_8 \oplus x_0x_1x_4x_7x_8 \oplus
x_2x_4x_7x_8 \oplus x_1x_2x_4x_7x_8 \oplus x_0x_1x_2x_4x_7x_8 \oplus
x_0x_3x_4x_7x_8 \oplus x_1x_3x_4x_7x_8 \oplus x_2x_3x_4x_7x_8 \oplus
x_0x_2x_3x_4x_7x_8 \oplus x_0x_1x_2x_3x_4x_7x_8 \oplus x_5x_7x_8
\oplus x_1x_5x_7x_8 \oplus x_0x_1x_5x_7x_8 \oplus x_0x_2x_5x_7x_8
\oplus x_1x_2x_5x_7x_8 \oplus x_3x_5x_7x_8 \oplus x_0x_3x_5x_7x_8
\oplus x_0x_1x_3x_5x_7x_8 \oplus x_2x_3x_5x_7x_8 \oplus
x_1x_2x_3x_5x_7x_8 \oplus x_0x_1x_2x_3x_5x_7x_8 \oplus x_0x_4x_5x_7x_8
\oplus x_1x_4x_5x_7x_8 \oplus x_2x_4x_5x_7x_8 \oplus
x_0x_2x_4x_5x_7x_8 \oplus x_0x_1x_2x_4x_5x_7x_8 \oplus x_3x_4x_5x_7x_8
\oplus x_1x_3x_4x_5x_7x_8 \oplus x_0x_1x_3x_4x_5x_7x_8 \oplus
x_0x_2x_3x_4x_5x_7x_8 \oplus x_1x_2x_3x_4x_5x_7x_8 \oplus x_6x_7x_8
\oplus x_0x_6x_7x_8 \oplus x_0x_1x_6x_7x_8 \oplus x_2x_6x_7x_8 \oplus
x_1x_2x_6x_7x_8 \oplus x_0x_1x_2x_6x_7x_8 \oplus x_0x_3x_6x_7x_8
\oplus x_1x_3x_6x_7x_8 \oplus x_2x_3x_6x_7x_8 \oplus
x_0x_2x_3x_6x_7x_8 \oplus x_0x_1x_2x_3x_6x_7x_8 \oplus x_4x_6x_7x_8
\oplus x_1x_4x_6x_7x_8 \oplus x_0x_1x_4x_6x_7x_8 \oplus
x_0x_2x_4x_6x_7x_8 \oplus x_1x_2x_4x_6x_7x_8 \oplus x_3x_4x_6x_7x_8
\oplus x_0x_3x_4x_6x_7x_8 \oplus x_0x_1x_3x_4x_6x_7x_8 \oplus
x_2x_3x_4x_6x_7x_8 \oplus x_1x_2x_3x_4x_6x_7x_8 \oplus
x_0x_1x_2x_3x_4x_6x_7x_8 \oplus x_0x_5x_6x_7x_8 \oplus x_1x_5x_6x_7x_8
\oplus x_2x_5x_6x_7x_8 \oplus x_0x_2x_5x_6x_7x_8 \oplus
x_0x_1x_2x_5x_6x_7x_8 \oplus x_3x_5x_6x_7x_8 \oplus x_1x_3x_5x_6x_7x_8
\oplus x_0x_1x_3x_5x_6x_7x_8 \oplus x_0x_2x_3x_5x_6x_7x_8 \oplus
x_1x_2x_3x_5x_6x_7x_8 \oplus x_4x_5x_6x_7x_8 \oplus x_0x_4x_5x_6x_7x_8
\oplus x_0x_1x_4x_5x_6x_7x_8 \oplus x_2x_4x_5x_6x_7x_8 \oplus
x_1x_2x_4x_5x_6x_7x_8 \oplus x_0x_1x_2x_4x_5x_6x_7x_8 \oplus
x_0x_3x_4x_5x_6x_7x_8 \oplus x_1x_3x_4x_5x_6x_7x_8 \oplus
x_2x_3x_4x_5x_6x_7x_8 \oplus x_0x_2x_3x_4x_5x_6x_7x_8 \oplus
x_0x_1x_2x_3x_4x_5x_6x_7x_8 \oplus x_9 \oplus x_1x_9 \oplus x_0x_1x_9
\oplus x_0x_2x_9 \oplus x_1x_2x_9 \oplus x_3x_9 \oplus x_0x_3x_9
\oplus x_0x_1x_3x_9 \oplus x_2x_3x_9 \oplus x_1x_2x_3x_9 \oplus
x_0x_1x_2x_3x_9 \oplus x_0x_4x_9 \oplus x_1x_4x_9 \oplus x_2x_4x_9
\oplus x_0x_2x_4x_9 \oplus x_0x_1x_2x_4x_9 \oplus x_3x_4x_9 \oplus
x_1x_3x_4x_9 \oplus x_0x_1x_3x_4x_9 \oplus x_0x_2x_3x_4x_9 \oplus
x_1x_2x_3x_4x_9 \oplus x_5x_9 \oplus x_0x_5x_9 \oplus x_0x_1x_5x_9
\oplus x_2x_5x_9 \oplus x_1x_2x_5x_9 \oplus x_0x_1x_2x_5x_9 \oplus
x_0x_3x_5x_9 \oplus x_1x_3x_5x_9 \oplus x_2x_3x_5x_9 \oplus
x_0x_2x_3x_5x_9 \oplus x_0x_1x_2x_3x_5x_9 \oplus x_4x_5x_9 \oplus
x_1x_4x_5x_9 \oplus x_0x_1x_4x_5x_9 \oplus x_0x_2x_4x_5x_9 \oplus
x_1x_2x_4x_5x_9 \oplus x_3x_4x_5x_9 \oplus x_0x_3x_4x_5x_9 \oplus
x_0x_1x_3x_4x_5x_9 \oplus x_2x_3x_4x_5x_9 \oplus x_1x_2x_3x_4x_5x_9
\oplus x_0x_1x_2x_3x_4x_5x_9 \oplus x_0x_6x_9 \oplus x_1x_6x_9 \oplus
x_2x_6x_9 \oplus x_0x_2x_6x_9 \oplus x_0x_1x_2x_6x_9 \oplus x_3x_6x_9
\oplus x_1x_3x_6x_9 \oplus x_0x_1x_3x_6x_9 \oplus x_0x_2x_3x_6x_9
\oplus x_1x_2x_3x_6x_9 \oplus x_4x_6x_9 \oplus x_0x_4x_6x_9 \oplus
x_0x_1x_4x_6x_9 \oplus x_2x_4x_6x_9 \oplus x_1x_2x_4x_6x_9 \oplus
x_0x_1x_2x_4x_6x_9 \oplus x_0x_3x_4x_6x_9 \oplus x_1x_3x_4x_6x_9
\oplus x_2x_3x_4x_6x_9 \oplus x_0x_2x_3x_4x_6x_9 \oplus
x_0x_1x_2x_3x_4x_6x_9 \oplus x_5x_6x_9 \oplus x_1x_5x_6x_9 \oplus
x_0x_1x_5x_6x_9 \oplus x_0x_2x_5x_6x_9 \oplus x_1x_2x_5x_6x_9 \oplus
x_3x_5x_6x_9 \oplus x_0x_3x_5x_6x_9 \oplus x_0x_1x_3x_5x_6x_9 \oplus
x_2x_3x_5x_6x_9 \oplus x_1x_2x_3x_5x_6x_9 \oplus x_0x_1x_2x_3x_5x_6x_9
\oplus x_0x_4x_5x_6x_9 \oplus x_1x_4x_5x_6x_9 \oplus x_2x_4x_5x_6x_9
\oplus x_0x_2x_4x_5x_6x_9 \oplus x_0x_1x_2x_4x_5x_6x_9 \oplus
x_3x_4x_5x_6x_9 \oplus x_1x_3x_4x_5x_6x_9 \oplus x_0x_1x_3x_4x_5x_6x_9
\oplus x_0x_2x_3x_4x_5x_6x_9 \oplus x_1x_2x_3x_4x_5x_6x_9 \oplus
x_7x_9 \oplus x_0x_7x_9 \oplus x_0x_1x_7x_9 \oplus x_2x_7x_9 \oplus
x_1x_2x_7x_9 \oplus x_0x_1x_2x_7x_9 \oplus x_0x_3x_7x_9 \oplus
x_1x_3x_7x_9 \oplus x_2x_3x_7x_9 \oplus x_0x_2x_3x_7x_9 \oplus
x_0x_1x_2x_3x_7x_9 \oplus x_4x_7x_9 \oplus x_1x_4x_7x_9 \oplus
x_0x_1x_4x_7x_9 \oplus x_0x_2x_4x_7x_9 \oplus x_1x_2x_4x_7x_9 \oplus
x_3x_4x_7x_9 \oplus x_0x_3x_4x_7x_9 \oplus x_0x_1x_3x_4x_7x_9 \oplus
x_2x_3x_4x_7x_9 \oplus x_1x_2x_3x_4x_7x_9 \oplus x_0x_1x_2x_3x_4x_7x_9
\oplus x_0x_5x_7x_9 \oplus x_1x_5x_7x_9 \oplus x_2x_5x_7x_9 \oplus
x_0x_2x_5x_7x_9 \oplus x_0x_1x_2x_5x_7x_9 \oplus x_3x_5x_7x_9 \oplus
x_1x_3x_5x_7x_9 \oplus x_0x_1x_3x_5x_7x_9 \oplus x_0x_2x_3x_5x_7x_9
\oplus x_1x_2x_3x_5x_7x_9 \oplus x_4x_5x_7x_9 \oplus x_0x_4x_5x_7x_9
\oplus x_0x_1x_4x_5x_7x_9 \oplus x_2x_4x_5x_7x_9 \oplus
x_1x_2x_4x_5x_7x_9 \oplus x_0x_1x_2x_4x_5x_7x_9 \oplus
x_0x_3x_4x_5x_7x_9 \oplus x_1x_3x_4x_5x_7x_9 \oplus x_2x_3x_4x_5x_7x_9
\oplus x_0x_2x_3x_4x_5x_7x_9 \oplus x_0x_1x_2x_3x_4x_5x_7x_9 \oplus
x_6x_7x_9 \oplus x_1x_6x_7x_9 \oplus x_0x_1x_6x_7x_9 \oplus
x_0x_2x_6x_7x_9 \oplus x_1x_2x_6x_7x_9 \oplus x_3x_6x_7x_9 \oplus
x_0x_3x_6x_7x_9 \oplus x_0x_1x_3x_6x_7x_9 \oplus x_2x_3x_6x_7x_9
\oplus x_1x_2x_3x_6x_7x_9 \oplus x_0x_1x_2x_3x_6x_7x_9 \oplus
x_0x_4x_6x_7x_9 \oplus x_1x_4x_6x_7x_9 \oplus x_2x_4x_6x_7x_9 \oplus
x_0x_2x_4x_6x_7x_9 \oplus x_0x_1x_2x_4x_6x_7x_9 \oplus x_3x_4x_6x_7x_9
\oplus x_1x_3x_4x_6x_7x_9 \oplus x_0x_1x_3x_4x_6x_7x_9 \oplus
x_0x_2x_3x_4x_6x_7x_9 \oplus x_1x_2x_3x_4x_6x_7x_9 \oplus x_5x_6x_7x_9
\oplus x_0x_5x_6x_7x_9 \oplus x_0x_1x_5x_6x_7x_9 \oplus
x_2x_5x_6x_7x_9 \oplus x_1x_2x_5x_6x_7x_9 \oplus x_0x_1x_2x_5x_6x_7x_9
\oplus x_0x_3x_5x_6x_7x_9 \oplus x_1x_3x_5x_6x_7x_9 \oplus
x_2x_3x_5x_6x_7x_9 \oplus x_0x_2x_3x_5x_6x_7x_9 \oplus
x_0x_1x_2x_3x_5x_6x_7x_9 \oplus x_4x_5x_6x_7x_9 \oplus
x_1x_4x_5x_6x_7x_9 \oplus x_0x_1x_4x_5x_6x_7x_9 \oplus
x_0x_2x_4x_5x_6x_7x_9 \oplus x_1x_2x_4x_5x_6x_7x_9 \oplus
x_3x_4x_5x_6x_7x_9 \oplus x_0x_3x_4x_5x_6x_7x_9 \oplus
x_0x_1x_3x_4x_5x_6x_7x_9 \oplus x_2x_3x_4x_5x_6x_7x_9 \oplus
x_1x_2x_3x_4x_5x_6x_7x_9 \oplus x_0x_1x_2x_3x_4x_5x_6x_7x_9 \oplus
x_0x_8x_9 \oplus x_1x_8x_9 \oplus x_2x_8x_9 \oplus x_0x_2x_8x_9 \oplus
x_0x_1x_2x_8x_9 \oplus x_3x_8x_9 \oplus x_1x_3x_8x_9 \oplus
x_0x_1x_3x_8x_9 \oplus x_0x_2x_3x_8x_9 \oplus x_1x_2x_3x_8x_9 \oplus
x_4x_8x_9 \oplus x_0x_4x_8x_9 \oplus x_0x_1x_4x_8x_9 \oplus
x_2x_4x_8x_9 \oplus x_1x_2x_4x_8x_9 \oplus x_0x_1x_2x_4x_8x_9 \oplus
x_0x_3x_4x_8x_9 \oplus x_1x_3x_4x_8x_9 \oplus x_2x_3x_4x_8x_9 \oplus
x_0x_2x_3x_4x_8x_9 \oplus x_0x_1x_2x_3x_4x_8x_9 \oplus x_5x_8x_9
\oplus x_1x_5x_8x_9 \oplus x_0x_1x_5x_8x_9 \oplus x_0x_2x_5x_8x_9
\oplus x_1x_2x_5x_8x_9 \oplus x_3x_5x_8x_9 \oplus x_0x_3x_5x_8x_9
\oplus x_0x_1x_3x_5x_8x_9 \oplus x_2x_3x_5x_8x_9 \oplus
x_1x_2x_3x_5x_8x_9 \oplus x_0x_1x_2x_3x_5x_8x_9 \oplus x_0x_4x_5x_8x_9
\oplus x_1x_4x_5x_8x_9 \oplus x_2x_4x_5x_8x_9 \oplus
x_0x_2x_4x_5x_8x_9 \oplus x_0x_1x_2x_4x_5x_8x_9 \oplus x_3x_4x_5x_8x_9
\oplus x_1x_3x_4x_5x_8x_9 \oplus x_0x_1x_3x_4x_5x_8x_9 \oplus
x_0x_2x_3x_4x_5x_8x_9 \oplus x_1x_2x_3x_4x_5x_8x_9 \oplus x_6x_8x_9
\oplus x_0x_6x_8x_9 \oplus x_0x_1x_6x_8x_9 \oplus x_2x_6x_8x_9 \oplus
x_1x_2x_6x_8x_9 \oplus x_0x_1x_2x_6x_8x_9 \oplus x_0x_3x_6x_8x_9
\oplus x_1x_3x_6x_8x_9 \oplus x_2x_3x_6x_8x_9 \oplus
x_0x_2x_3x_6x_8x_9 \oplus x_0x_1x_2x_3x_6x_8x_9 \oplus x_4x_6x_8x_9
\oplus x_1x_4x_6x_8x_9 \oplus x_0x_1x_4x_6x_8x_9 \oplus
x_0x_2x_4x_6x_8x_9 \oplus x_1x_2x_4x_6x_8x_9 \oplus x_3x_4x_6x_8x_9
\oplus x_0x_3x_4x_6x_8x_9 \oplus x_0x_1x_3x_4x_6x_8x_9 \oplus
x_2x_3x_4x_6x_8x_9 \oplus x_1x_2x_3x_4x_6x_8x_9 \oplus
x_0x_1x_2x_3x_4x_6x_8x_9 \oplus x_0x_5x_6x_8x_9 \oplus x_1x_5x_6x_8x_9
\oplus x_2x_5x_6x_8x_9 \oplus x_0x_2x_5x_6x_8x_9 \oplus
x_0x_1x_2x_5x_6x_8x_9 \oplus x_3x_5x_6x_8x_9 \oplus x_1x_3x_5x_6x_8x_9
\oplus x_0x_1x_3x_5x_6x_8x_9 \oplus x_0x_2x_3x_5x_6x_8x_9 \oplus
x_1x_2x_3x_5x_6x_8x_9 \oplus x_4x_5x_6x_8x_9 \oplus x_0x_4x_5x_6x_8x_9
\oplus x_0x_1x_4x_5x_6x_8x_9 \oplus x_2x_4x_5x_6x_8x_9 \oplus
x_1x_2x_4x_5x_6x_8x_9 \oplus x_0x_1x_2x_4x_5x_6x_8x_9 \oplus
x_0x_3x_4x_5x_6x_8x_9 \oplus x_1x_3x_4x_5x_6x_8x_9 \oplus
x_2x_3x_4x_5x_6x_8x_9 \oplus x_0x_2x_3x_4x_5x_6x_8x_9 \oplus
x_0x_1x_2x_3x_4x_5x_6x_8x_9 \oplus x_7x_8x_9 \oplus x_1x_7x_8x_9
\oplus x_0x_1x_7x_8x_9 \oplus x_0x_2x_7x_8x_9 \oplus x_1x_2x_7x_8x_9
\oplus x_3x_7x_8x_9 \oplus x_0x_3x_7x_8x_9 \oplus x_0x_1x_3x_7x_8x_9
\oplus x_2x_3x_7x_8x_9 \oplus x_1x_2x_3x_7x_8x_9 \oplus
x_0x_1x_2x_3x_7x_8x_9 \oplus x_0x_4x_7x_8x_9 \oplus x_1x_4x_7x_8x_9
\oplus x_2x_4x_7x_8x_9 \oplus x_0x_2x_4x_7x_8x_9 \oplus
x_0x_1x_2x_4x_7x_8x_9 \oplus x_3x_4x_7x_8x_9 \oplus x_1x_3x_4x_7x_8x_9
\oplus x_0x_1x_3x_4x_7x_8x_9 \oplus x_0x_2x_3x_4x_7x_8x_9 \oplus
x_1x_2x_3x_4x_7x_8x_9 \oplus x_5x_7x_8x_9 \oplus x_0x_5x_7x_8x_9
\oplus x_0x_1x_5x_7x_8x_9 \oplus x_2x_5x_7x_8x_9 \oplus
x_1x_2x_5x_7x_8x_9 \oplus x_0x_1x_2x_5x_7x_8x_9 \oplus
x_0x_3x_5x_7x_8x_9 \oplus x_1x_3x_5x_7x_8x_9 \oplus x_2x_3x_5x_7x_8x_9
\oplus x_0x_2x_3x_5x_7x_8x_9 \oplus x_0x_1x_2x_3x_5x_7x_8x_9 \oplus
x_4x_5x_7x_8x_9 \oplus x_1x_4x_5x_7x_8x_9 \oplus x_0x_1x_4x_5x_7x_8x_9
\oplus x_0x_2x_4x_5x_7x_8x_9 \oplus x_1x_2x_4x_5x_7x_8x_9 \oplus
x_3x_4x_5x_7x_8x_9 \oplus x_0x_3x_4x_5x_7x_8x_9 \oplus
x_0x_1x_3x_4x_5x_7x_8x_9 \oplus x_2x_3x_4x_5x_7x_8x_9 \oplus
x_1x_2x_3x_4x_5x_7x_8x_9 \oplus x_0x_1x_2x_3x_4x_5x_7x_8x_9 \oplus
x_0x_6x_7x_8x_9 \oplus x_1x_6x_7x_8x_9 \oplus x_2x_6x_7x_8x_9 \oplus
x_0x_2x_6x_7x_8x_9 \oplus x_0x_1x_2x_6x_7x_8x_9 \oplus x_3x_6x_7x_8x_9
\oplus x_1x_3x_6x_7x_8x_9 \oplus x_0x_1x_3x_6x_7x_8x_9 \oplus
x_0x_2x_3x_6x_7x_8x_9 \oplus x_1x_2x_3x_6x_7x_8x_9 \oplus
x_4x_6x_7x_8x_9 \oplus x_0x_4x_6x_7x_8x_9 \oplus x_0x_1x_4x_6x_7x_8x_9
\oplus x_2x_4x_6x_7x_8x_9 \oplus x_1x_2x_4x_6x_7x_8x_9 \oplus
x_0x_1x_2x_4x_6x_7x_8x_9 \oplus x_0x_3x_4x_6x_7x_8x_9 \oplus
x_1x_3x_4x_6x_7x_8x_9 \oplus x_2x_3x_4x_6x_7x_8x_9 \oplus
x_0x_2x_3x_4x_6x_7x_8x_9 \oplus x_0x_1x_2x_3x_4x_6x_7x_8x_9 \oplus
x_5x_6x_7x_8x_9 \oplus x_1x_5x_6x_7x_8x_9 \oplus x_0x_1x_5x_6x_7x_8x_9
\oplus x_0x_2x_5x_6x_7x_8x_9 \oplus x_1x_2x_5x_6x_7x_8x_9 \oplus
x_3x_5x_6x_7x_8x_9 \oplus x_0x_3x_5x_6x_7x_8x_9 \oplus
x_0x_1x_3x_5x_6x_7x_8x_9 \oplus x_2x_3x_5x_6x_7x_8x_9 \oplus
x_1x_2x_3x_5x_6x_7x_8x_9 \oplus x_0x_1x_2x_3x_5x_6x_7x_8x_9 \oplus
x_0x_4x_5x_6x_7x_8x_9 \oplus x_1x_4x_5x_6x_7x_8x_9 \oplus
x_2x_4x_5x_6x_7x_8x_9 \oplus x_0x_2x_4x_5x_6x_7x_8x_9 \oplus
x_0x_1x_2x_4x_5x_6x_7x_8x_9 \oplus x_3x_4x_5x_6x_7x_8x_9 \oplus
x_1x_3x_4x_5x_6x_7x_8x_9 \oplus x_0x_1x_3x_4x_5x_6x_7x_8x_9 \oplus
x_0x_2x_3x_4x_5x_6x_7x_8x_9 \oplus x_1x_2x_3x_4x_5x_6x_7x_8x_9 = 1$

\bigskip

$x_1 \oplus x_0x_1 \oplus x_0x_2 \oplus x_1x_2 \oplus x_3 \oplus
x_0x_3 \oplus x_0x_1x_3 \oplus x_2x_3 \oplus x_1x_2x_3 \oplus
x_0x_1x_2x_3 \oplus x_0x_4 \oplus x_1x_4 \oplus x_2x_4 \oplus
x_0x_2x_4 \oplus x_0x_1x_2x_4 \oplus x_3x_4 \oplus x_1x_3x_4 \oplus
x_0x_1x_3x_4 \oplus x_0x_2x_3x_4 \oplus x_1x_2x_3x_4 \oplus x_5 \oplus
x_0x_5 \oplus x_0x_1x_5 \oplus x_2x_5 \oplus x_1x_2x_5 \oplus
x_0x_1x_2x_5 \oplus x_0x_3x_5 \oplus x_1x_3x_5 \oplus x_2x_3x_5 \oplus
x_0x_2x_3x_5 \oplus x_0x_1x_2x_3x_5 \oplus x_4x_5 \oplus x_1x_4x_5
\oplus x_0x_1x_4x_5 \oplus x_0x_2x_4x_5 \oplus x_1x_2x_4x_5 \oplus
x_3x_4x_5 \oplus x_0x_3x_4x_5 \oplus x_0x_1x_3x_4x_5 \oplus
x_2x_3x_4x_5 \oplus x_1x_2x_3x_4x_5 \oplus x_0x_1x_2x_3x_4x_5 \oplus
x_0x_6 \oplus x_1x_6 \oplus x_2x_6 \oplus x_0x_2x_6 \oplus
x_0x_1x_2x_6 \oplus x_3x_6 \oplus x_1x_3x_6 \oplus x_0x_1x_3x_6 \oplus
x_0x_2x_3x_6 \oplus x_1x_2x_3x_6 \oplus x_4x_6 \oplus x_0x_4x_6 \oplus
x_0x_1x_4x_6 \oplus x_2x_4x_6 \oplus x_1x_2x_4x_6 \oplus
x_0x_1x_2x_4x_6 \oplus x_0x_3x_4x_6 \oplus x_1x_3x_4x_6 \oplus
x_2x_3x_4x_6 \oplus x_0x_2x_3x_4x_6 \oplus x_0x_1x_2x_3x_4x_6 \oplus
x_5x_6 \oplus x_1x_5x_6 \oplus x_0x_1x_5x_6 \oplus x_0x_2x_5x_6 \oplus
x_1x_2x_5x_6 \oplus x_3x_5x_6 \oplus x_0x_3x_5x_6 \oplus
x_0x_1x_3x_5x_6 \oplus x_2x_3x_5x_6 \oplus x_1x_2x_3x_5x_6 \oplus
x_0x_1x_2x_3x_5x_6 \oplus x_0x_4x_5x_6 \oplus x_1x_4x_5x_6 \oplus
x_2x_4x_5x_6 \oplus x_0x_2x_4x_5x_6 \oplus x_0x_1x_2x_4x_5x_6 \oplus
x_3x_4x_5x_6 \oplus x_1x_3x_4x_5x_6 \oplus x_0x_1x_3x_4x_5x_6 \oplus
x_0x_2x_3x_4x_5x_6 \oplus x_1x_2x_3x_4x_5x_6 \oplus x_7 \oplus x_0x_7
\oplus x_0x_1x_7 \oplus x_2x_7 \oplus x_1x_2x_7 \oplus x_0x_1x_2x_7
\oplus x_0x_3x_7 \oplus x_1x_3x_7 \oplus x_2x_3x_7 \oplus x_0x_2x_3x_7
\oplus x_0x_1x_2x_3x_7 \oplus x_4x_7 \oplus x_1x_4x_7 \oplus
x_0x_1x_4x_7 \oplus x_0x_2x_4x_7 \oplus x_1x_2x_4x_7 \oplus x_3x_4x_7
\oplus x_0x_3x_4x_7 \oplus x_0x_1x_3x_4x_7 \oplus x_2x_3x_4x_7 \oplus
x_1x_2x_3x_4x_7 \oplus x_0x_1x_2x_3x_4x_7 \oplus x_0x_5x_7 \oplus
x_1x_5x_7 \oplus x_2x_5x_7 \oplus x_0x_2x_5x_7 \oplus x_0x_1x_2x_5x_7
\oplus x_3x_5x_7 \oplus x_1x_3x_5x_7 \oplus x_0x_1x_3x_5x_7 \oplus
x_0x_2x_3x_5x_7 \oplus x_1x_2x_3x_5x_7 \oplus x_4x_5x_7 \oplus
x_0x_4x_5x_7 \oplus x_0x_1x_4x_5x_7 \oplus x_2x_4x_5x_7 \oplus
x_1x_2x_4x_5x_7 \oplus x_0x_1x_2x_4x_5x_7 \oplus x_0x_3x_4x_5x_7
\oplus x_1x_3x_4x_5x_7 \oplus x_2x_3x_4x_5x_7 \oplus
x_0x_2x_3x_4x_5x_7 \oplus x_0x_1x_2x_3x_4x_5x_7 \oplus x_6x_7 \oplus
x_1x_6x_7 \oplus x_0x_1x_6x_7 \oplus x_0x_2x_6x_7 \oplus x_1x_2x_6x_7
\oplus x_3x_6x_7 \oplus x_0x_3x_6x_7 \oplus x_0x_1x_3x_6x_7 \oplus
x_2x_3x_6x_7 \oplus x_1x_2x_3x_6x_7 \oplus x_0x_1x_2x_3x_6x_7 \oplus
x_0x_4x_6x_7 \oplus x_1x_4x_6x_7 \oplus x_2x_4x_6x_7 \oplus
x_0x_2x_4x_6x_7 \oplus x_0x_1x_2x_4x_6x_7 \oplus x_3x_4x_6x_7 \oplus
x_1x_3x_4x_6x_7 \oplus x_0x_1x_3x_4x_6x_7 \oplus x_0x_2x_3x_4x_6x_7
\oplus x_1x_2x_3x_4x_6x_7 \oplus x_5x_6x_7 \oplus x_0x_5x_6x_7 \oplus
x_0x_1x_5x_6x_7 \oplus x_2x_5x_6x_7 \oplus x_1x_2x_5x_6x_7 \oplus
x_0x_1x_2x_5x_6x_7 \oplus x_0x_3x_5x_6x_7 \oplus x_1x_3x_5x_6x_7
\oplus x_2x_3x_5x_6x_7 \oplus x_0x_2x_3x_5x_6x_7 \oplus
x_0x_1x_2x_3x_5x_6x_7 \oplus x_4x_5x_6x_7 \oplus x_1x_4x_5x_6x_7
\oplus x_0x_1x_4x_5x_6x_7 \oplus x_0x_2x_4x_5x_6x_7 \oplus
x_1x_2x_4x_5x_6x_7 \oplus x_3x_4x_5x_6x_7 \oplus x_0x_3x_4x_5x_6x_7
\oplus x_0x_1x_3x_4x_5x_6x_7 \oplus x_2x_3x_4x_5x_6x_7 \oplus
x_1x_2x_3x_4x_5x_6x_7 \oplus x_0x_1x_2x_3x_4x_5x_6x_7 \oplus x_0x_8
\oplus x_1x_8 \oplus x_2x_8 \oplus x_0x_2x_8 \oplus x_0x_1x_2x_8
\oplus x_3x_8 \oplus x_1x_3x_8 \oplus x_0x_1x_3x_8 \oplus x_0x_2x_3x_8
\oplus x_1x_2x_3x_8 \oplus x_4x_8 \oplus x_0x_4x_8 \oplus x_0x_1x_4x_8
\oplus x_2x_4x_8 \oplus x_1x_2x_4x_8 \oplus x_0x_1x_2x_4x_8 \oplus
x_0x_3x_4x_8 \oplus x_1x_3x_4x_8 \oplus x_2x_3x_4x_8 \oplus
x_0x_2x_3x_4x_8 \oplus x_0x_1x_2x_3x_4x_8 \oplus x_5x_8 \oplus
x_1x_5x_8 \oplus x_0x_1x_5x_8 \oplus x_0x_2x_5x_8 \oplus x_1x_2x_5x_8
\oplus x_3x_5x_8 \oplus x_0x_3x_5x_8 \oplus x_0x_1x_3x_5x_8 \oplus
x_2x_3x_5x_8 \oplus x_1x_2x_3x_5x_8 \oplus x_0x_1x_2x_3x_5x_8 \oplus
x_0x_4x_5x_8 \oplus x_1x_4x_5x_8 \oplus x_2x_4x_5x_8 \oplus
x_0x_2x_4x_5x_8 \oplus x_0x_1x_2x_4x_5x_8 \oplus x_3x_4x_5x_8 \oplus
x_1x_3x_4x_5x_8 \oplus x_0x_1x_3x_4x_5x_8 \oplus x_0x_2x_3x_4x_5x_8
\oplus x_1x_2x_3x_4x_5x_8 \oplus x_6x_8 \oplus x_0x_6x_8 \oplus
x_0x_1x_6x_8 \oplus x_2x_6x_8 \oplus x_1x_2x_6x_8 \oplus
x_0x_1x_2x_6x_8 \oplus x_0x_3x_6x_8 \oplus x_1x_3x_6x_8 \oplus
x_2x_3x_6x_8 \oplus x_0x_2x_3x_6x_8 \oplus x_0x_1x_2x_3x_6x_8 \oplus
x_4x_6x_8 \oplus x_1x_4x_6x_8 \oplus x_0x_1x_4x_6x_8 \oplus
x_0x_2x_4x_6x_8 \oplus x_1x_2x_4x_6x_8 \oplus x_3x_4x_6x_8 \oplus
x_0x_3x_4x_6x_8 \oplus x_0x_1x_3x_4x_6x_8 \oplus x_2x_3x_4x_6x_8
\oplus x_1x_2x_3x_4x_6x_8 \oplus x_0x_1x_2x_3x_4x_6x_8 \oplus
x_0x_5x_6x_8 \oplus x_1x_5x_6x_8 \oplus x_2x_5x_6x_8 \oplus
x_0x_2x_5x_6x_8 \oplus x_0x_1x_2x_5x_6x_8 \oplus x_3x_5x_6x_8 \oplus
x_1x_3x_5x_6x_8 \oplus x_0x_1x_3x_5x_6x_8 \oplus x_0x_2x_3x_5x_6x_8
\oplus x_1x_2x_3x_5x_6x_8 \oplus x_4x_5x_6x_8 \oplus x_0x_4x_5x_6x_8
\oplus x_0x_1x_4x_5x_6x_8 \oplus x_2x_4x_5x_6x_8 \oplus
x_1x_2x_4x_5x_6x_8 \oplus x_0x_1x_2x_4x_5x_6x_8 \oplus
x_0x_3x_4x_5x_6x_8 \oplus x_1x_3x_4x_5x_6x_8 \oplus x_2x_3x_4x_5x_6x_8
\oplus x_0x_2x_3x_4x_5x_6x_8 \oplus x_0x_1x_2x_3x_4x_5x_6x_8 \oplus
x_7x_8 \oplus x_1x_7x_8 \oplus x_0x_1x_7x_8 \oplus x_0x_2x_7x_8 \oplus
x_1x_2x_7x_8 \oplus x_3x_7x_8 \oplus x_0x_3x_7x_8 \oplus
x_0x_1x_3x_7x_8 \oplus x_2x_3x_7x_8 \oplus x_1x_2x_3x_7x_8 \oplus
x_0x_1x_2x_3x_7x_8 \oplus x_0x_4x_7x_8 \oplus x_1x_4x_7x_8 \oplus
x_2x_4x_7x_8 \oplus x_0x_2x_4x_7x_8 \oplus x_0x_1x_2x_4x_7x_8 \oplus
x_3x_4x_7x_8 \oplus x_1x_3x_4x_7x_8 \oplus x_0x_1x_3x_4x_7x_8 \oplus
x_0x_2x_3x_4x_7x_8 \oplus x_1x_2x_3x_4x_7x_8 \oplus x_5x_7x_8 \oplus
x_0x_5x_7x_8 \oplus x_0x_1x_5x_7x_8 \oplus x_2x_5x_7x_8 \oplus
x_1x_2x_5x_7x_8 \oplus x_0x_1x_2x_5x_7x_8 \oplus x_0x_3x_5x_7x_8
\oplus x_1x_3x_5x_7x_8 \oplus x_2x_3x_5x_7x_8 \oplus
x_0x_2x_3x_5x_7x_8 \oplus x_0x_1x_2x_3x_5x_7x_8 \oplus x_4x_5x_7x_8
\oplus x_1x_4x_5x_7x_8 \oplus x_0x_1x_4x_5x_7x_8 \oplus
x_0x_2x_4x_5x_7x_8 \oplus x_1x_2x_4x_5x_7x_8 \oplus x_3x_4x_5x_7x_8
\oplus x_0x_3x_4x_5x_7x_8 \oplus x_0x_1x_3x_4x_5x_7x_8 \oplus
x_2x_3x_4x_5x_7x_8 \oplus x_1x_2x_3x_4x_5x_7x_8 \oplus
x_0x_1x_2x_3x_4x_5x_7x_8 \oplus x_0x_6x_7x_8 \oplus x_1x_6x_7x_8
\oplus x_2x_6x_7x_8 \oplus x_0x_2x_6x_7x_8 \oplus x_0x_1x_2x_6x_7x_8
\oplus x_3x_6x_7x_8 \oplus x_1x_3x_6x_7x_8 \oplus x_0x_1x_3x_6x_7x_8
\oplus x_0x_2x_3x_6x_7x_8 \oplus x_1x_2x_3x_6x_7x_8 \oplus
x_4x_6x_7x_8 \oplus x_0x_4x_6x_7x_8 \oplus x_0x_1x_4x_6x_7x_8 \oplus
x_2x_4x_6x_7x_8 \oplus x_1x_2x_4x_6x_7x_8 \oplus x_0x_1x_2x_4x_6x_7x_8
\oplus x_0x_3x_4x_6x_7x_8 \oplus x_1x_3x_4x_6x_7x_8 \oplus
x_2x_3x_4x_6x_7x_8 \oplus x_0x_2x_3x_4x_6x_7x_8 \oplus
x_0x_1x_2x_3x_4x_6x_7x_8 \oplus x_5x_6x_7x_8 \oplus x_1x_5x_6x_7x_8
\oplus x_0x_1x_5x_6x_7x_8 \oplus x_0x_2x_5x_6x_7x_8 \oplus
x_1x_2x_5x_6x_7x_8 \oplus x_3x_5x_6x_7x_8 \oplus x_0x_3x_5x_6x_7x_8
\oplus x_0x_1x_3x_5x_6x_7x_8 \oplus x_2x_3x_5x_6x_7x_8 \oplus
x_1x_2x_3x_5x_6x_7x_8 \oplus x_0x_1x_2x_3x_5x_6x_7x_8 \oplus
x_0x_4x_5x_6x_7x_8 \oplus x_1x_4x_5x_6x_7x_8 \oplus x_2x_4x_5x_6x_7x_8
\oplus x_0x_2x_4x_5x_6x_7x_8 \oplus x_0x_1x_2x_4x_5x_6x_7x_8 \oplus
x_3x_4x_5x_6x_7x_8 \oplus x_1x_3x_4x_5x_6x_7x_8 \oplus
x_0x_1x_3x_4x_5x_6x_7x_8 \oplus x_0x_2x_3x_4x_5x_6x_7x_8 \oplus
x_1x_2x_3x_4x_5x_6x_7x_8 \oplus x_9 \oplus x_0x_9 \oplus x_0x_1x_9
\oplus x_2x_9 \oplus x_1x_2x_9 \oplus x_0x_1x_2x_9 \oplus x_0x_3x_9
\oplus x_1x_3x_9 \oplus x_2x_3x_9 \oplus x_0x_2x_3x_9 \oplus
x_0x_1x_2x_3x_9 \oplus x_4x_9 \oplus x_1x_4x_9 \oplus x_0x_1x_4x_9
\oplus x_0x_2x_4x_9 \oplus x_1x_2x_4x_9 \oplus x_3x_4x_9 \oplus
x_0x_3x_4x_9 \oplus x_0x_1x_3x_4x_9 \oplus x_2x_3x_4x_9 \oplus
x_1x_2x_3x_4x_9 \oplus x_0x_1x_2x_3x_4x_9 \oplus x_0x_5x_9 \oplus
x_1x_5x_9 \oplus x_2x_5x_9 \oplus x_0x_2x_5x_9 \oplus x_0x_1x_2x_5x_9
\oplus x_3x_5x_9 \oplus x_1x_3x_5x_9 \oplus x_0x_1x_3x_5x_9 \oplus
x_0x_2x_3x_5x_9 \oplus x_1x_2x_3x_5x_9 \oplus x_4x_5x_9 \oplus
x_0x_4x_5x_9 \oplus x_0x_1x_4x_5x_9 \oplus x_2x_4x_5x_9 \oplus
x_1x_2x_4x_5x_9 \oplus x_0x_1x_2x_4x_5x_9 \oplus x_0x_3x_4x_5x_9
\oplus x_1x_3x_4x_5x_9 \oplus x_2x_3x_4x_5x_9 \oplus
x_0x_2x_3x_4x_5x_9 \oplus x_0x_1x_2x_3x_4x_5x_9 \oplus x_6x_9 \oplus
x_1x_6x_9 \oplus x_0x_1x_6x_9 \oplus x_0x_2x_6x_9 \oplus x_1x_2x_6x_9
\oplus x_3x_6x_9 \oplus x_0x_3x_6x_9 \oplus x_0x_1x_3x_6x_9 \oplus
x_2x_3x_6x_9 \oplus x_1x_2x_3x_6x_9 \oplus x_0x_1x_2x_3x_6x_9 \oplus
x_0x_4x_6x_9 \oplus x_1x_4x_6x_9 \oplus x_2x_4x_6x_9 \oplus
x_0x_2x_4x_6x_9 \oplus x_0x_1x_2x_4x_6x_9 \oplus x_3x_4x_6x_9 \oplus
x_1x_3x_4x_6x_9 \oplus x_0x_1x_3x_4x_6x_9 \oplus x_0x_2x_3x_4x_6x_9
\oplus x_1x_2x_3x_4x_6x_9 \oplus x_5x_6x_9 \oplus x_0x_5x_6x_9 \oplus
x_0x_1x_5x_6x_9 \oplus x_2x_5x_6x_9 \oplus x_1x_2x_5x_6x_9 \oplus
x_0x_1x_2x_5x_6x_9 \oplus x_0x_3x_5x_6x_9 \oplus x_1x_3x_5x_6x_9
\oplus x_2x_3x_5x_6x_9 \oplus x_0x_2x_3x_5x_6x_9 \oplus
x_0x_1x_2x_3x_5x_6x_9 \oplus x_4x_5x_6x_9 \oplus x_1x_4x_5x_6x_9
\oplus x_0x_1x_4x_5x_6x_9 \oplus x_0x_2x_4x_5x_6x_9 \oplus
x_1x_2x_4x_5x_6x_9 \oplus x_3x_4x_5x_6x_9 \oplus x_0x_3x_4x_5x_6x_9
\oplus x_0x_1x_3x_4x_5x_6x_9 \oplus x_2x_3x_4x_5x_6x_9 \oplus
x_1x_2x_3x_4x_5x_6x_9 \oplus x_0x_1x_2x_3x_4x_5x_6x_9 \oplus x_0x_7x_9
\oplus x_1x_7x_9 \oplus x_2x_7x_9 \oplus x_0x_2x_7x_9 \oplus
x_0x_1x_2x_7x_9 \oplus x_3x_7x_9 \oplus x_1x_3x_7x_9 \oplus
x_0x_1x_3x_7x_9 \oplus x_0x_2x_3x_7x_9 \oplus x_1x_2x_3x_7x_9 \oplus
x_4x_7x_9 \oplus x_0x_4x_7x_9 \oplus x_0x_1x_4x_7x_9 \oplus
x_2x_4x_7x_9 \oplus x_1x_2x_4x_7x_9 \oplus x_0x_1x_2x_4x_7x_9 \oplus
x_0x_3x_4x_7x_9 \oplus x_1x_3x_4x_7x_9 \oplus x_2x_3x_4x_7x_9 \oplus
x_0x_2x_3x_4x_7x_9 \oplus x_0x_1x_2x_3x_4x_7x_9 \oplus x_5x_7x_9
\oplus x_1x_5x_7x_9 \oplus x_0x_1x_5x_7x_9 \oplus x_0x_2x_5x_7x_9
\oplus x_1x_2x_5x_7x_9 \oplus x_3x_5x_7x_9 \oplus x_0x_3x_5x_7x_9
\oplus x_0x_1x_3x_5x_7x_9 \oplus x_2x_3x_5x_7x_9 \oplus
x_1x_2x_3x_5x_7x_9 \oplus x_0x_1x_2x_3x_5x_7x_9 \oplus x_0x_4x_5x_7x_9
\oplus x_1x_4x_5x_7x_9 \oplus x_2x_4x_5x_7x_9 \oplus
x_0x_2x_4x_5x_7x_9 \oplus x_0x_1x_2x_4x_5x_7x_9 \oplus x_3x_4x_5x_7x_9
\oplus x_1x_3x_4x_5x_7x_9 \oplus x_0x_1x_3x_4x_5x_7x_9 \oplus
x_0x_2x_3x_4x_5x_7x_9 \oplus x_1x_2x_3x_4x_5x_7x_9 \oplus x_6x_7x_9
\oplus x_0x_6x_7x_9 \oplus x_0x_1x_6x_7x_9 \oplus x_2x_6x_7x_9 \oplus
x_1x_2x_6x_7x_9 \oplus x_0x_1x_2x_6x_7x_9 \oplus x_0x_3x_6x_7x_9
\oplus x_1x_3x_6x_7x_9 \oplus x_2x_3x_6x_7x_9 \oplus
x_0x_2x_3x_6x_7x_9 \oplus x_0x_1x_2x_3x_6x_7x_9 \oplus x_4x_6x_7x_9
\oplus x_1x_4x_6x_7x_9 \oplus x_0x_1x_4x_6x_7x_9 \oplus
x_0x_2x_4x_6x_7x_9 \oplus x_1x_2x_4x_6x_7x_9 \oplus x_3x_4x_6x_7x_9
\oplus x_0x_3x_4x_6x_7x_9 \oplus x_0x_1x_3x_4x_6x_7x_9 \oplus
x_2x_3x_4x_6x_7x_9 \oplus x_1x_2x_3x_4x_6x_7x_9 \oplus
x_0x_1x_2x_3x_4x_6x_7x_9 \oplus x_0x_5x_6x_7x_9 \oplus x_1x_5x_6x_7x_9
\oplus x_2x_5x_6x_7x_9 \oplus x_0x_2x_5x_6x_7x_9 \oplus
x_0x_1x_2x_5x_6x_7x_9 \oplus x_3x_5x_6x_7x_9 \oplus x_1x_3x_5x_6x_7x_9
\oplus x_0x_1x_3x_5x_6x_7x_9 \oplus x_0x_2x_3x_5x_6x_7x_9 \oplus
x_1x_2x_3x_5x_6x_7x_9 \oplus x_4x_5x_6x_7x_9 \oplus x_0x_4x_5x_6x_7x_9
\oplus x_0x_1x_4x_5x_6x_7x_9 \oplus x_2x_4x_5x_6x_7x_9 \oplus
x_1x_2x_4x_5x_6x_7x_9 \oplus x_0x_1x_2x_4x_5x_6x_7x_9 \oplus
x_0x_3x_4x_5x_6x_7x_9 \oplus x_1x_3x_4x_5x_6x_7x_9 \oplus
x_2x_3x_4x_5x_6x_7x_9 \oplus x_0x_2x_3x_4x_5x_6x_7x_9 \oplus
x_0x_1x_2x_3x_4x_5x_6x_7x_9 \oplus x_8x_9 \oplus x_1x_8x_9 \oplus
x_0x_1x_8x_9 \oplus x_0x_2x_8x_9 \oplus x_1x_2x_8x_9 \oplus x_3x_8x_9
\oplus x_0x_3x_8x_9 \oplus x_0x_1x_3x_8x_9 \oplus x_2x_3x_8x_9 \oplus
x_1x_2x_3x_8x_9 \oplus x_0x_1x_2x_3x_8x_9 \oplus x_0x_4x_8x_9 \oplus
x_1x_4x_8x_9 \oplus x_2x_4x_8x_9 \oplus x_0x_2x_4x_8x_9 \oplus
x_0x_1x_2x_4x_8x_9 \oplus x_3x_4x_8x_9 \oplus x_1x_3x_4x_8x_9 \oplus
x_0x_1x_3x_4x_8x_9 \oplus x_0x_2x_3x_4x_8x_9 \oplus x_1x_2x_3x_4x_8x_9
\oplus x_5x_8x_9 \oplus x_0x_5x_8x_9 \oplus x_0x_1x_5x_8x_9 \oplus
x_2x_5x_8x_9 \oplus x_1x_2x_5x_8x_9 \oplus x_0x_1x_2x_5x_8x_9 \oplus
x_0x_3x_5x_8x_9 \oplus x_1x_3x_5x_8x_9 \oplus x_2x_3x_5x_8x_9 \oplus
x_0x_2x_3x_5x_8x_9 \oplus x_0x_1x_2x_3x_5x_8x_9 \oplus x_4x_5x_8x_9
\oplus x_1x_4x_5x_8x_9 \oplus x_0x_1x_4x_5x_8x_9 \oplus
x_0x_2x_4x_5x_8x_9 \oplus x_1x_2x_4x_5x_8x_9 \oplus x_3x_4x_5x_8x_9
\oplus x_0x_3x_4x_5x_8x_9 \oplus x_0x_1x_3x_4x_5x_8x_9 \oplus
x_2x_3x_4x_5x_8x_9 \oplus x_1x_2x_3x_4x_5x_8x_9 \oplus
x_0x_1x_2x_3x_4x_5x_8x_9 \oplus x_0x_6x_8x_9 \oplus x_1x_6x_8x_9
\oplus x_2x_6x_8x_9 \oplus x_0x_2x_6x_8x_9 \oplus x_0x_1x_2x_6x_8x_9
\oplus x_3x_6x_8x_9 \oplus x_1x_3x_6x_8x_9 \oplus x_0x_1x_3x_6x_8x_9
\oplus x_0x_2x_3x_6x_8x_9 \oplus x_1x_2x_3x_6x_8x_9 \oplus
x_4x_6x_8x_9 \oplus x_0x_4x_6x_8x_9 \oplus x_0x_1x_4x_6x_8x_9 \oplus
x_2x_4x_6x_8x_9 \oplus x_1x_2x_4x_6x_8x_9 \oplus x_0x_1x_2x_4x_6x_8x_9
\oplus x_0x_3x_4x_6x_8x_9 \oplus x_1x_3x_4x_6x_8x_9 \oplus
x_2x_3x_4x_6x_8x_9 \oplus x_0x_2x_3x_4x_6x_8x_9 \oplus
x_0x_1x_2x_3x_4x_6x_8x_9 \oplus x_5x_6x_8x_9 \oplus x_1x_5x_6x_8x_9
\oplus x_0x_1x_5x_6x_8x_9 \oplus x_0x_2x_5x_6x_8x_9 \oplus
x_1x_2x_5x_6x_8x_9 \oplus x_3x_5x_6x_8x_9 \oplus x_0x_3x_5x_6x_8x_9
\oplus x_0x_1x_3x_5x_6x_8x_9 \oplus x_2x_3x_5x_6x_8x_9 \oplus
x_1x_2x_3x_5x_6x_8x_9 \oplus x_0x_1x_2x_3x_5x_6x_8x_9 \oplus
x_0x_4x_5x_6x_8x_9 \oplus x_1x_4x_5x_6x_8x_9 \oplus x_2x_4x_5x_6x_8x_9
\oplus x_0x_2x_4x_5x_6x_8x_9 \oplus x_0x_1x_2x_4x_5x_6x_8x_9 \oplus
x_3x_4x_5x_6x_8x_9 \oplus x_1x_3x_4x_5x_6x_8x_9 \oplus
x_0x_1x_3x_4x_5x_6x_8x_9 \oplus x_0x_2x_3x_4x_5x_6x_8x_9 \oplus
x_1x_2x_3x_4x_5x_6x_8x_9 \oplus x_7x_8x_9 \oplus x_0x_7x_8x_9 \oplus
x_0x_1x_7x_8x_9 \oplus x_2x_7x_8x_9 \oplus x_1x_2x_7x_8x_9 \oplus
x_0x_1x_2x_7x_8x_9 \oplus x_0x_3x_7x_8x_9 \oplus x_1x_3x_7x_8x_9
\oplus x_2x_3x_7x_8x_9 \oplus x_0x_2x_3x_7x_8x_9 \oplus
x_0x_1x_2x_3x_7x_8x_9 \oplus x_4x_7x_8x_9 \oplus x_1x_4x_7x_8x_9
\oplus x_0x_1x_4x_7x_8x_9 \oplus x_0x_2x_4x_7x_8x_9 \oplus
x_1x_2x_4x_7x_8x_9 \oplus x_3x_4x_7x_8x_9 \oplus x_0x_3x_4x_7x_8x_9
\oplus x_0x_1x_3x_4x_7x_8x_9 \oplus x_2x_3x_4x_7x_8x_9 \oplus
x_1x_2x_3x_4x_7x_8x_9 \oplus x_0x_1x_2x_3x_4x_7x_8x_9 \oplus
x_0x_5x_7x_8x_9 \oplus x_1x_5x_7x_8x_9 \oplus x_2x_5x_7x_8x_9 \oplus
x_0x_2x_5x_7x_8x_9 \oplus x_0x_1x_2x_5x_7x_8x_9 \oplus x_3x_5x_7x_8x_9
\oplus x_1x_3x_5x_7x_8x_9 \oplus x_0x_1x_3x_5x_7x_8x_9 \oplus
x_0x_2x_3x_5x_7x_8x_9 \oplus x_1x_2x_3x_5x_7x_8x_9 \oplus
x_4x_5x_7x_8x_9 \oplus x_0x_4x_5x_7x_8x_9 \oplus x_0x_1x_4x_5x_7x_8x_9
\oplus x_2x_4x_5x_7x_8x_9 \oplus x_1x_2x_4x_5x_7x_8x_9 \oplus
x_0x_1x_2x_4x_5x_7x_8x_9 \oplus x_0x_3x_4x_5x_7x_8x_9 \oplus
x_1x_3x_4x_5x_7x_8x_9 \oplus x_2x_3x_4x_5x_7x_8x_9 \oplus
x_0x_2x_3x_4x_5x_7x_8x_9 \oplus x_0x_1x_2x_3x_4x_5x_7x_8x_9 \oplus
x_6x_7x_8x_9 \oplus x_1x_6x_7x_8x_9 \oplus x_0x_1x_6x_7x_8x_9 \oplus
x_0x_2x_6x_7x_8x_9 \oplus x_1x_2x_6x_7x_8x_9 \oplus x_3x_6x_7x_8x_9
\oplus x_0x_3x_6x_7x_8x_9 \oplus x_0x_1x_3x_6x_7x_8x_9 \oplus
x_2x_3x_6x_7x_8x_9 \oplus x_1x_2x_3x_6x_7x_8x_9 \oplus
x_0x_1x_2x_3x_6x_7x_8x_9 \oplus x_0x_4x_6x_7x_8x_9 \oplus
x_1x_4x_6x_7x_8x_9 \oplus x_2x_4x_6x_7x_8x_9 \oplus
x_0x_2x_4x_6x_7x_8x_9 \oplus x_0x_1x_2x_4x_6x_7x_8x_9 \oplus
x_3x_4x_6x_7x_8x_9 \oplus x_1x_3x_4x_6x_7x_8x_9 \oplus
x_0x_1x_3x_4x_6x_7x_8x_9 \oplus x_0x_2x_3x_4x_6x_7x_8x_9 \oplus
x_1x_2x_3x_4x_6x_7x_8x_9 \oplus x_5x_6x_7x_8x_9 \oplus
x_0x_5x_6x_7x_8x_9 \oplus x_0x_1x_5x_6x_7x_8x_9 \oplus
x_2x_5x_6x_7x_8x_9 \oplus x_1x_2x_5x_6x_7x_8x_9 \oplus
x_0x_1x_2x_5x_6x_7x_8x_9 \oplus x_0x_3x_5x_6x_7x_8x_9 \oplus
x_1x_3x_5x_6x_7x_8x_9 \oplus x_2x_3x_5x_6x_7x_8x_9 \oplus
x_0x_2x_3x_5x_6x_7x_8x_9 \oplus x_0x_1x_2x_3x_5x_6x_7x_8x_9 \oplus
x_4x_5x_6x_7x_8x_9 \oplus x_1x_4x_5x_6x_7x_8x_9 \oplus
x_0x_1x_4x_5x_6x_7x_8x_9 \oplus x_0x_2x_4x_5x_6x_7x_8x_9 \oplus
x_1x_2x_4x_5x_6x_7x_8x_9 \oplus x_3x_4x_5x_6x_7x_8x_9 \oplus
x_0x_3x_4x_5x_6x_7x_8x_9 \oplus x_0x_1x_3x_4x_5x_6x_7x_8x_9 \oplus
x_2x_3x_4x_5x_6x_7x_8x_9 \oplus x_1x_2x_3x_4x_5x_6x_7x_8x_9 \oplus
x_0x_1x_2x_3x_4x_5x_6x_7x_8x_9 = 0$

\bigskip

$x_0 \oplus x_0x_1 \oplus x_2 \oplus x_1x_2 \oplus x_0x_1x_2 \oplus
x_0x_3 \oplus x_1x_3 \oplus x_2x_3 \oplus x_0x_2x_3 \oplus
x_0x_1x_2x_3 \oplus x_4 \oplus x_1x_4 \oplus x_0x_1x_4 \oplus
x_0x_2x_4 \oplus x_1x_2x_4 \oplus x_3x_4 \oplus x_0x_3x_4 \oplus
x_0x_1x_3x_4 \oplus x_2x_3x_4 \oplus x_1x_2x_3x_4 \oplus
x_0x_1x_2x_3x_4 \oplus x_0x_5 \oplus x_1x_5 \oplus x_2x_5 \oplus
x_0x_2x_5 \oplus x_0x_1x_2x_5 \oplus x_3x_5 \oplus x_1x_3x_5 \oplus
x_0x_1x_3x_5 \oplus x_0x_2x_3x_5 \oplus x_1x_2x_3x_5 \oplus x_4x_5
\oplus x_0x_4x_5 \oplus x_0x_1x_4x_5 \oplus x_2x_4x_5 \oplus
x_1x_2x_4x_5 \oplus x_0x_1x_2x_4x_5 \oplus x_0x_3x_4x_5 \oplus
x_1x_3x_4x_5 \oplus x_2x_3x_4x_5 \oplus x_0x_2x_3x_4x_5 \oplus
x_0x_1x_2x_3x_4x_5 \oplus x_6 \oplus x_1x_6 \oplus x_0x_1x_6 \oplus
x_0x_2x_6 \oplus x_1x_2x_6 \oplus x_3x_6 \oplus x_0x_3x_6 \oplus
x_0x_1x_3x_6 \oplus x_2x_3x_6 \oplus x_1x_2x_3x_6 \oplus
x_0x_1x_2x_3x_6 \oplus x_0x_4x_6 \oplus x_1x_4x_6 \oplus x_2x_4x_6
\oplus x_0x_2x_4x_6 \oplus x_0x_1x_2x_4x_6 \oplus x_3x_4x_6 \oplus
x_1x_3x_4x_6 \oplus x_0x_1x_3x_4x_6 \oplus x_0x_2x_3x_4x_6 \oplus
x_1x_2x_3x_4x_6 \oplus x_5x_6 \oplus x_0x_5x_6 \oplus x_0x_1x_5x_6
\oplus x_2x_5x_6 \oplus x_1x_2x_5x_6 \oplus x_0x_1x_2x_5x_6 \oplus
x_0x_3x_5x_6 \oplus x_1x_3x_5x_6 \oplus x_2x_3x_5x_6 \oplus
x_0x_2x_3x_5x_6 \oplus x_0x_1x_2x_3x_5x_6 \oplus x_4x_5x_6 \oplus
x_1x_4x_5x_6 \oplus x_0x_1x_4x_5x_6 \oplus x_0x_2x_4x_5x_6 \oplus
x_1x_2x_4x_5x_6 \oplus x_3x_4x_5x_6 \oplus x_0x_3x_4x_5x_6 \oplus
x_0x_1x_3x_4x_5x_6 \oplus x_2x_3x_4x_5x_6 \oplus x_1x_2x_3x_4x_5x_6
\oplus x_0x_1x_2x_3x_4x_5x_6 \oplus x_0x_7 \oplus x_1x_7 \oplus x_2x_7
\oplus x_0x_2x_7 \oplus x_0x_1x_2x_7 \oplus x_3x_7 \oplus x_1x_3x_7
\oplus x_0x_1x_3x_7 \oplus x_0x_2x_3x_7 \oplus x_1x_2x_3x_7 \oplus
x_4x_7 \oplus x_0x_4x_7 \oplus x_0x_1x_4x_7 \oplus x_2x_4x_7 \oplus
x_1x_2x_4x_7 \oplus x_0x_1x_2x_4x_7 \oplus x_0x_3x_4x_7 \oplus
x_1x_3x_4x_7 \oplus x_2x_3x_4x_7 \oplus x_0x_2x_3x_4x_7 \oplus
x_0x_1x_2x_3x_4x_7 \oplus x_5x_7 \oplus x_1x_5x_7 \oplus x_0x_1x_5x_7
\oplus x_0x_2x_5x_7 \oplus x_1x_2x_5x_7 \oplus x_3x_5x_7 \oplus
x_0x_3x_5x_7 \oplus x_0x_1x_3x_5x_7 \oplus x_2x_3x_5x_7 \oplus
x_1x_2x_3x_5x_7 \oplus x_0x_1x_2x_3x_5x_7 \oplus x_0x_4x_5x_7 \oplus
x_1x_4x_5x_7 \oplus x_2x_4x_5x_7 \oplus x_0x_2x_4x_5x_7 \oplus
x_0x_1x_2x_4x_5x_7 \oplus x_3x_4x_5x_7 \oplus x_1x_3x_4x_5x_7 \oplus
x_0x_1x_3x_4x_5x_7 \oplus x_0x_2x_3x_4x_5x_7 \oplus x_1x_2x_3x_4x_5x_7
\oplus x_6x_7 \oplus x_0x_6x_7 \oplus x_0x_1x_6x_7 \oplus x_2x_6x_7
\oplus x_1x_2x_6x_7 \oplus x_0x_1x_2x_6x_7 \oplus x_0x_3x_6x_7 \oplus
x_1x_3x_6x_7 \oplus x_2x_3x_6x_7 \oplus x_0x_2x_3x_6x_7 \oplus
x_0x_1x_2x_3x_6x_7 \oplus x_4x_6x_7 \oplus x_1x_4x_6x_7 \oplus
x_0x_1x_4x_6x_7 \oplus x_0x_2x_4x_6x_7 \oplus x_1x_2x_4x_6x_7 \oplus
x_3x_4x_6x_7 \oplus x_0x_3x_4x_6x_7 \oplus x_0x_1x_3x_4x_6x_7 \oplus
x_2x_3x_4x_6x_7 \oplus x_1x_2x_3x_4x_6x_7 \oplus x_0x_1x_2x_3x_4x_6x_7
\oplus x_0x_5x_6x_7 \oplus x_1x_5x_6x_7 \oplus x_2x_5x_6x_7 \oplus
x_0x_2x_5x_6x_7 \oplus x_0x_1x_2x_5x_6x_7 \oplus x_3x_5x_6x_7 \oplus
x_1x_3x_5x_6x_7 \oplus x_0x_1x_3x_5x_6x_7 \oplus x_0x_2x_3x_5x_6x_7
\oplus x_1x_2x_3x_5x_6x_7 \oplus x_4x_5x_6x_7 \oplus x_0x_4x_5x_6x_7
\oplus x_0x_1x_4x_5x_6x_7 \oplus x_2x_4x_5x_6x_7 \oplus
x_1x_2x_4x_5x_6x_7 \oplus x_0x_1x_2x_4x_5x_6x_7 \oplus
x_0x_3x_4x_5x_6x_7 \oplus x_1x_3x_4x_5x_6x_7 \oplus x_2x_3x_4x_5x_6x_7
\oplus x_0x_2x_3x_4x_5x_6x_7 \oplus x_0x_1x_2x_3x_4x_5x_6x_7 \oplus
x_8 \oplus x_1x_8 \oplus x_0x_1x_8 \oplus x_0x_2x_8 \oplus x_1x_2x_8
\oplus x_3x_8 \oplus x_0x_3x_8 \oplus x_0x_1x_3x_8 \oplus x_2x_3x_8
\oplus x_1x_2x_3x_8 \oplus x_0x_1x_2x_3x_8 \oplus x_0x_4x_8 \oplus
x_1x_4x_8 \oplus x_2x_4x_8 \oplus x_0x_2x_4x_8 \oplus x_0x_1x_2x_4x_8
\oplus x_3x_4x_8 \oplus x_1x_3x_4x_8 \oplus x_0x_1x_3x_4x_8 \oplus
x_0x_2x_3x_4x_8 \oplus x_1x_2x_3x_4x_8 \oplus x_5x_8 \oplus x_0x_5x_8
\oplus x_0x_1x_5x_8 \oplus x_2x_5x_8 \oplus x_1x_2x_5x_8 \oplus
x_0x_1x_2x_5x_8 \oplus x_0x_3x_5x_8 \oplus x_1x_3x_5x_8 \oplus
x_2x_3x_5x_8 \oplus x_0x_2x_3x_5x_8 \oplus x_0x_1x_2x_3x_5x_8 \oplus
x_4x_5x_8 \oplus x_1x_4x_5x_8 \oplus x_0x_1x_4x_5x_8 \oplus
x_0x_2x_4x_5x_8 \oplus x_1x_2x_4x_5x_8 \oplus x_3x_4x_5x_8 \oplus
x_0x_3x_4x_5x_8 \oplus x_0x_1x_3x_4x_5x_8 \oplus x_2x_3x_4x_5x_8
\oplus x_1x_2x_3x_4x_5x_8 \oplus x_0x_1x_2x_3x_4x_5x_8 \oplus
x_0x_6x_8 \oplus x_1x_6x_8 \oplus x_2x_6x_8 \oplus x_0x_2x_6x_8 \oplus
x_0x_1x_2x_6x_8 \oplus x_3x_6x_8 \oplus x_1x_3x_6x_8 \oplus
x_0x_1x_3x_6x_8 \oplus x_0x_2x_3x_6x_8 \oplus x_1x_2x_3x_6x_8 \oplus
x_4x_6x_8 \oplus x_0x_4x_6x_8 \oplus x_0x_1x_4x_6x_8 \oplus
x_2x_4x_6x_8 \oplus x_1x_2x_4x_6x_8 \oplus x_0x_1x_2x_4x_6x_8 \oplus
x_0x_3x_4x_6x_8 \oplus x_1x_3x_4x_6x_8 \oplus x_2x_3x_4x_6x_8 \oplus
x_0x_2x_3x_4x_6x_8 \oplus x_0x_1x_2x_3x_4x_6x_8 \oplus x_5x_6x_8
\oplus x_1x_5x_6x_8 \oplus x_0x_1x_5x_6x_8 \oplus x_0x_2x_5x_6x_8
\oplus x_1x_2x_5x_6x_8 \oplus x_3x_5x_6x_8 \oplus x_0x_3x_5x_6x_8
\oplus x_0x_1x_3x_5x_6x_8 \oplus x_2x_3x_5x_6x_8 \oplus
x_1x_2x_3x_5x_6x_8 \oplus x_0x_1x_2x_3x_5x_6x_8 \oplus x_0x_4x_5x_6x_8
\oplus x_1x_4x_5x_6x_8 \oplus x_2x_4x_5x_6x_8 \oplus
x_0x_2x_4x_5x_6x_8 \oplus x_0x_1x_2x_4x_5x_6x_8 \oplus x_3x_4x_5x_6x_8
\oplus x_1x_3x_4x_5x_6x_8 \oplus x_0x_1x_3x_4x_5x_6x_8 \oplus
x_0x_2x_3x_4x_5x_6x_8 \oplus x_1x_2x_3x_4x_5x_6x_8 \oplus x_7x_8
\oplus x_0x_7x_8 \oplus x_0x_1x_7x_8 \oplus x_2x_7x_8 \oplus
x_1x_2x_7x_8 \oplus x_0x_1x_2x_7x_8 \oplus x_0x_3x_7x_8 \oplus
x_1x_3x_7x_8 \oplus x_2x_3x_7x_8 \oplus x_0x_2x_3x_7x_8 \oplus
x_0x_1x_2x_3x_7x_8 \oplus x_4x_7x_8 \oplus x_1x_4x_7x_8 \oplus
x_0x_1x_4x_7x_8 \oplus x_0x_2x_4x_7x_8 \oplus x_1x_2x_4x_7x_8 \oplus
x_3x_4x_7x_8 \oplus x_0x_3x_4x_7x_8 \oplus x_0x_1x_3x_4x_7x_8 \oplus
x_2x_3x_4x_7x_8 \oplus x_1x_2x_3x_4x_7x_8 \oplus x_0x_1x_2x_3x_4x_7x_8
\oplus x_0x_5x_7x_8 \oplus x_1x_5x_7x_8 \oplus x_2x_5x_7x_8 \oplus
x_0x_2x_5x_7x_8 \oplus x_0x_1x_2x_5x_7x_8 \oplus x_3x_5x_7x_8 \oplus
x_1x_3x_5x_7x_8 \oplus x_0x_1x_3x_5x_7x_8 \oplus x_0x_2x_3x_5x_7x_8
\oplus x_1x_2x_3x_5x_7x_8 \oplus x_4x_5x_7x_8 \oplus x_0x_4x_5x_7x_8
\oplus x_0x_1x_4x_5x_7x_8 \oplus x_2x_4x_5x_7x_8 \oplus
x_1x_2x_4x_5x_7x_8 \oplus x_0x_1x_2x_4x_5x_7x_8 \oplus
x_0x_3x_4x_5x_7x_8 \oplus x_1x_3x_4x_5x_7x_8 \oplus x_2x_3x_4x_5x_7x_8
\oplus x_0x_2x_3x_4x_5x_7x_8 \oplus x_0x_1x_2x_3x_4x_5x_7x_8 \oplus
x_6x_7x_8 \oplus x_1x_6x_7x_8 \oplus x_0x_1x_6x_7x_8 \oplus
x_0x_2x_6x_7x_8 \oplus x_1x_2x_6x_7x_8 \oplus x_3x_6x_7x_8 \oplus
x_0x_3x_6x_7x_8 \oplus x_0x_1x_3x_6x_7x_8 \oplus x_2x_3x_6x_7x_8
\oplus x_1x_2x_3x_6x_7x_8 \oplus x_0x_1x_2x_3x_6x_7x_8 \oplus
x_0x_4x_6x_7x_8 \oplus x_1x_4x_6x_7x_8 \oplus x_2x_4x_6x_7x_8 \oplus
x_0x_2x_4x_6x_7x_8 \oplus x_0x_1x_2x_4x_6x_7x_8 \oplus x_3x_4x_6x_7x_8
\oplus x_1x_3x_4x_6x_7x_8 \oplus x_0x_1x_3x_4x_6x_7x_8 \oplus
x_0x_2x_3x_4x_6x_7x_8 \oplus x_1x_2x_3x_4x_6x_7x_8 \oplus x_5x_6x_7x_8
\oplus x_0x_5x_6x_7x_8 \oplus x_0x_1x_5x_6x_7x_8 \oplus
x_2x_5x_6x_7x_8 \oplus x_1x_2x_5x_6x_7x_8 \oplus x_0x_1x_2x_5x_6x_7x_8
\oplus x_0x_3x_5x_6x_7x_8 \oplus x_1x_3x_5x_6x_7x_8 \oplus
x_2x_3x_5x_6x_7x_8 \oplus x_0x_2x_3x_5x_6x_7x_8 \oplus
x_0x_1x_2x_3x_5x_6x_7x_8 \oplus x_4x_5x_6x_7x_8 \oplus
x_1x_4x_5x_6x_7x_8 \oplus x_0x_1x_4x_5x_6x_7x_8 \oplus
x_0x_2x_4x_5x_6x_7x_8 \oplus x_1x_2x_4x_5x_6x_7x_8 \oplus
x_3x_4x_5x_6x_7x_8 \oplus x_0x_3x_4x_5x_6x_7x_8 \oplus
x_0x_1x_3x_4x_5x_6x_7x_8 \oplus x_2x_3x_4x_5x_6x_7x_8 \oplus
x_1x_2x_3x_4x_5x_6x_7x_8 \oplus x_0x_1x_2x_3x_4x_5x_6x_7x_8 \oplus
x_0x_9 \oplus x_1x_9 \oplus x_2x_9 \oplus x_0x_2x_9 \oplus
x_0x_1x_2x_9 \oplus x_3x_9 \oplus x_1x_3x_9 \oplus x_0x_1x_3x_9 \oplus
x_0x_2x_3x_9 \oplus x_1x_2x_3x_9 \oplus x_4x_9 \oplus x_0x_4x_9 \oplus
x_0x_1x_4x_9 \oplus x_2x_4x_9 \oplus x_1x_2x_4x_9 \oplus
x_0x_1x_2x_4x_9 \oplus x_0x_3x_4x_9 \oplus x_1x_3x_4x_9 \oplus
x_2x_3x_4x_9 \oplus x_0x_2x_3x_4x_9 \oplus x_0x_1x_2x_3x_4x_9 \oplus
x_5x_9 \oplus x_1x_5x_9 \oplus x_0x_1x_5x_9 \oplus x_0x_2x_5x_9 \oplus
x_1x_2x_5x_9 \oplus x_3x_5x_9 \oplus x_0x_3x_5x_9 \oplus
x_0x_1x_3x_5x_9 \oplus x_2x_3x_5x_9 \oplus x_1x_2x_3x_5x_9 \oplus
x_0x_1x_2x_3x_5x_9 \oplus x_0x_4x_5x_9 \oplus x_1x_4x_5x_9 \oplus
x_2x_4x_5x_9 \oplus x_0x_2x_4x_5x_9 \oplus x_0x_1x_2x_4x_5x_9 \oplus
x_3x_4x_5x_9 \oplus x_1x_3x_4x_5x_9 \oplus x_0x_1x_3x_4x_5x_9 \oplus
x_0x_2x_3x_4x_5x_9 \oplus x_1x_2x_3x_4x_5x_9 \oplus x_6x_9 \oplus
x_0x_6x_9 \oplus x_0x_1x_6x_9 \oplus x_2x_6x_9 \oplus x_1x_2x_6x_9
\oplus x_0x_1x_2x_6x_9 \oplus x_0x_3x_6x_9 \oplus x_1x_3x_6x_9 \oplus
x_2x_3x_6x_9 \oplus x_0x_2x_3x_6x_9 \oplus x_0x_1x_2x_3x_6x_9 \oplus
x_4x_6x_9 \oplus x_1x_4x_6x_9 \oplus x_0x_1x_4x_6x_9 \oplus
x_0x_2x_4x_6x_9 \oplus x_1x_2x_4x_6x_9 \oplus x_3x_4x_6x_9 \oplus
x_0x_3x_4x_6x_9 \oplus x_0x_1x_3x_4x_6x_9 \oplus x_2x_3x_4x_6x_9
\oplus x_1x_2x_3x_4x_6x_9 \oplus x_0x_1x_2x_3x_4x_6x_9 \oplus
x_0x_5x_6x_9 \oplus x_1x_5x_6x_9 \oplus x_2x_5x_6x_9 \oplus
x_0x_2x_5x_6x_9 \oplus x_0x_1x_2x_5x_6x_9 \oplus x_3x_5x_6x_9 \oplus
x_1x_3x_5x_6x_9 \oplus x_0x_1x_3x_5x_6x_9 \oplus x_0x_2x_3x_5x_6x_9
\oplus x_1x_2x_3x_5x_6x_9 \oplus x_4x_5x_6x_9 \oplus x_0x_4x_5x_6x_9
\oplus x_0x_1x_4x_5x_6x_9 \oplus x_2x_4x_5x_6x_9 \oplus
x_1x_2x_4x_5x_6x_9 \oplus x_0x_1x_2x_4x_5x_6x_9 \oplus
x_0x_3x_4x_5x_6x_9 \oplus x_1x_3x_4x_5x_6x_9 \oplus x_2x_3x_4x_5x_6x_9
\oplus x_0x_2x_3x_4x_5x_6x_9 \oplus x_0x_1x_2x_3x_4x_5x_6x_9 \oplus
x_7x_9 \oplus x_1x_7x_9 \oplus x_0x_1x_7x_9 \oplus x_0x_2x_7x_9 \oplus
x_1x_2x_7x_9 \oplus x_3x_7x_9 \oplus x_0x_3x_7x_9 \oplus
x_0x_1x_3x_7x_9 \oplus x_2x_3x_7x_9 \oplus x_1x_2x_3x_7x_9 \oplus
x_0x_1x_2x_3x_7x_9 \oplus x_0x_4x_7x_9 \oplus x_1x_4x_7x_9 \oplus
x_2x_4x_7x_9 \oplus x_0x_2x_4x_7x_9 \oplus x_0x_1x_2x_4x_7x_9 \oplus
x_3x_4x_7x_9 \oplus x_1x_3x_4x_7x_9 \oplus x_0x_1x_3x_4x_7x_9 \oplus
x_0x_2x_3x_4x_7x_9 \oplus x_1x_2x_3x_4x_7x_9 \oplus x_5x_7x_9 \oplus
x_0x_5x_7x_9 \oplus x_0x_1x_5x_7x_9 \oplus x_2x_5x_7x_9 \oplus
x_1x_2x_5x_7x_9 \oplus x_0x_1x_2x_5x_7x_9 \oplus x_0x_3x_5x_7x_9
\oplus x_1x_3x_5x_7x_9 \oplus x_2x_3x_5x_7x_9 \oplus
x_0x_2x_3x_5x_7x_9 \oplus x_0x_1x_2x_3x_5x_7x_9 \oplus x_4x_5x_7x_9
\oplus x_1x_4x_5x_7x_9 \oplus x_0x_1x_4x_5x_7x_9 \oplus
x_0x_2x_4x_5x_7x_9 \oplus x_1x_2x_4x_5x_7x_9 \oplus x_3x_4x_5x_7x_9
\oplus x_0x_3x_4x_5x_7x_9 \oplus x_0x_1x_3x_4x_5x_7x_9 \oplus
x_2x_3x_4x_5x_7x_9 \oplus x_1x_2x_3x_4x_5x_7x_9 \oplus
x_0x_1x_2x_3x_4x_5x_7x_9 \oplus x_0x_6x_7x_9 \oplus x_1x_6x_7x_9
\oplus x_2x_6x_7x_9 \oplus x_0x_2x_6x_7x_9 \oplus x_0x_1x_2x_6x_7x_9
\oplus x_3x_6x_7x_9 \oplus x_1x_3x_6x_7x_9 \oplus x_0x_1x_3x_6x_7x_9
\oplus x_0x_2x_3x_6x_7x_9 \oplus x_1x_2x_3x_6x_7x_9 \oplus
x_4x_6x_7x_9 \oplus x_0x_4x_6x_7x_9 \oplus x_0x_1x_4x_6x_7x_9 \oplus
x_2x_4x_6x_7x_9 \oplus x_1x_2x_4x_6x_7x_9 \oplus x_0x_1x_2x_4x_6x_7x_9
\oplus x_0x_3x_4x_6x_7x_9 \oplus x_1x_3x_4x_6x_7x_9 \oplus
x_2x_3x_4x_6x_7x_9 \oplus x_0x_2x_3x_4x_6x_7x_9 \oplus
x_0x_1x_2x_3x_4x_6x_7x_9 \oplus x_5x_6x_7x_9 \oplus x_1x_5x_6x_7x_9
\oplus x_0x_1x_5x_6x_7x_9 \oplus x_0x_2x_5x_6x_7x_9 \oplus
x_1x_2x_5x_6x_7x_9 \oplus x_3x_5x_6x_7x_9 \oplus x_0x_3x_5x_6x_7x_9
\oplus x_0x_1x_3x_5x_6x_7x_9 \oplus x_2x_3x_5x_6x_7x_9 \oplus
x_1x_2x_3x_5x_6x_7x_9 \oplus x_0x_1x_2x_3x_5x_6x_7x_9 \oplus
x_0x_4x_5x_6x_7x_9 \oplus x_1x_4x_5x_6x_7x_9 \oplus x_2x_4x_5x_6x_7x_9
\oplus x_0x_2x_4x_5x_6x_7x_9 \oplus x_0x_1x_2x_4x_5x_6x_7x_9 \oplus
x_3x_4x_5x_6x_7x_9 \oplus x_1x_3x_4x_5x_6x_7x_9 \oplus
x_0x_1x_3x_4x_5x_6x_7x_9 \oplus x_0x_2x_3x_4x_5x_6x_7x_9 \oplus
x_1x_2x_3x_4x_5x_6x_7x_9 \oplus x_8x_9 \oplus x_0x_8x_9 \oplus
x_0x_1x_8x_9 \oplus x_2x_8x_9 \oplus x_1x_2x_8x_9 \oplus
x_0x_1x_2x_8x_9 \oplus x_0x_3x_8x_9 \oplus x_1x_3x_8x_9 \oplus
x_2x_3x_8x_9 \oplus x_0x_2x_3x_8x_9 \oplus x_0x_1x_2x_3x_8x_9 \oplus
x_4x_8x_9 \oplus x_1x_4x_8x_9 \oplus x_0x_1x_4x_8x_9 \oplus
x_0x_2x_4x_8x_9 \oplus x_1x_2x_4x_8x_9 \oplus x_3x_4x_8x_9 \oplus
x_0x_3x_4x_8x_9 \oplus x_0x_1x_3x_4x_8x_9 \oplus x_2x_3x_4x_8x_9
\oplus x_1x_2x_3x_4x_8x_9 \oplus x_0x_1x_2x_3x_4x_8x_9 \oplus
x_0x_5x_8x_9 \oplus x_1x_5x_8x_9 \oplus x_2x_5x_8x_9 \oplus
x_0x_2x_5x_8x_9 \oplus x_0x_1x_2x_5x_8x_9 \oplus x_3x_5x_8x_9 \oplus
x_1x_3x_5x_8x_9 \oplus x_0x_1x_3x_5x_8x_9 \oplus x_0x_2x_3x_5x_8x_9
\oplus x_1x_2x_3x_5x_8x_9 \oplus x_4x_5x_8x_9 \oplus x_0x_4x_5x_8x_9
\oplus x_0x_1x_4x_5x_8x_9 \oplus x_2x_4x_5x_8x_9 \oplus
x_1x_2x_4x_5x_8x_9 \oplus x_0x_1x_2x_4x_5x_8x_9 \oplus
x_0x_3x_4x_5x_8x_9 \oplus x_1x_3x_4x_5x_8x_9 \oplus x_2x_3x_4x_5x_8x_9
\oplus x_0x_2x_3x_4x_5x_8x_9 \oplus x_0x_1x_2x_3x_4x_5x_8x_9 \oplus
x_6x_8x_9 \oplus x_1x_6x_8x_9 \oplus x_0x_1x_6x_8x_9 \oplus
x_0x_2x_6x_8x_9 \oplus x_1x_2x_6x_8x_9 \oplus x_3x_6x_8x_9 \oplus
x_0x_3x_6x_8x_9 \oplus x_0x_1x_3x_6x_8x_9 \oplus x_2x_3x_6x_8x_9
\oplus x_1x_2x_3x_6x_8x_9 \oplus x_0x_1x_2x_3x_6x_8x_9 \oplus
x_0x_4x_6x_8x_9 \oplus x_1x_4x_6x_8x_9 \oplus x_2x_4x_6x_8x_9 \oplus
x_0x_2x_4x_6x_8x_9 \oplus x_0x_1x_2x_4x_6x_8x_9 \oplus x_3x_4x_6x_8x_9
\oplus x_1x_3x_4x_6x_8x_9 \oplus x_0x_1x_3x_4x_6x_8x_9 \oplus
x_0x_2x_3x_4x_6x_8x_9 \oplus x_1x_2x_3x_4x_6x_8x_9 \oplus x_5x_6x_8x_9
\oplus x_0x_5x_6x_8x_9 \oplus x_0x_1x_5x_6x_8x_9 \oplus
x_2x_5x_6x_8x_9 \oplus x_1x_2x_5x_6x_8x_9 \oplus x_0x_1x_2x_5x_6x_8x_9
\oplus x_0x_3x_5x_6x_8x_9 \oplus x_1x_3x_5x_6x_8x_9 \oplus
x_2x_3x_5x_6x_8x_9 \oplus x_0x_2x_3x_5x_6x_8x_9 \oplus
x_0x_1x_2x_3x_5x_6x_8x_9 \oplus x_4x_5x_6x_8x_9 \oplus
x_1x_4x_5x_6x_8x_9 \oplus x_0x_1x_4x_5x_6x_8x_9 \oplus
x_0x_2x_4x_5x_6x_8x_9 \oplus x_1x_2x_4x_5x_6x_8x_9 \oplus
x_3x_4x_5x_6x_8x_9 \oplus x_0x_3x_4x_5x_6x_8x_9 \oplus
x_0x_1x_3x_4x_5x_6x_8x_9 \oplus x_2x_3x_4x_5x_6x_8x_9 \oplus
x_1x_2x_3x_4x_5x_6x_8x_9 \oplus x_0x_1x_2x_3x_4x_5x_6x_8x_9 \oplus
x_0x_7x_8x_9 \oplus x_1x_7x_8x_9 \oplus x_2x_7x_8x_9 \oplus
x_0x_2x_7x_8x_9 \oplus x_0x_1x_2x_7x_8x_9 \oplus x_3x_7x_8x_9 \oplus
x_1x_3x_7x_8x_9 \oplus x_0x_1x_3x_7x_8x_9 \oplus x_0x_2x_3x_7x_8x_9
\oplus x_1x_2x_3x_7x_8x_9 \oplus x_4x_7x_8x_9 \oplus x_0x_4x_7x_8x_9
\oplus x_0x_1x_4x_7x_8x_9 \oplus x_2x_4x_7x_8x_9 \oplus
x_1x_2x_4x_7x_8x_9 \oplus x_0x_1x_2x_4x_7x_8x_9 \oplus
x_0x_3x_4x_7x_8x_9 \oplus x_1x_3x_4x_7x_8x_9 \oplus x_2x_3x_4x_7x_8x_9
\oplus x_0x_2x_3x_4x_7x_8x_9 \oplus x_0x_1x_2x_3x_4x_7x_8x_9 \oplus
x_5x_7x_8x_9 \oplus x_1x_5x_7x_8x_9 \oplus x_0x_1x_5x_7x_8x_9 \oplus
x_0x_2x_5x_7x_8x_9 \oplus x_1x_2x_5x_7x_8x_9 \oplus x_3x_5x_7x_8x_9
\oplus x_0x_3x_5x_7x_8x_9 \oplus x_0x_1x_3x_5x_7x_8x_9 \oplus
x_2x_3x_5x_7x_8x_9 \oplus x_1x_2x_3x_5x_7x_8x_9 \oplus
x_0x_1x_2x_3x_5x_7x_8x_9 \oplus x_0x_4x_5x_7x_8x_9 \oplus
x_1x_4x_5x_7x_8x_9 \oplus x_2x_4x_5x_7x_8x_9 \oplus
x_0x_2x_4x_5x_7x_8x_9 \oplus x_0x_1x_2x_4x_5x_7x_8x_9 \oplus
x_3x_4x_5x_7x_8x_9 \oplus x_1x_3x_4x_5x_7x_8x_9 \oplus
x_0x_1x_3x_4x_5x_7x_8x_9 \oplus x_0x_2x_3x_4x_5x_7x_8x_9 \oplus
x_1x_2x_3x_4x_5x_7x_8x_9 \oplus x_6x_7x_8x_9 \oplus x_0x_6x_7x_8x_9
\oplus x_0x_1x_6x_7x_8x_9 \oplus x_2x_6x_7x_8x_9 \oplus
x_1x_2x_6x_7x_8x_9 \oplus x_0x_1x_2x_6x_7x_8x_9 \oplus
x_0x_3x_6x_7x_8x_9 \oplus x_1x_3x_6x_7x_8x_9 \oplus x_2x_3x_6x_7x_8x_9
\oplus x_0x_2x_3x_6x_7x_8x_9 \oplus x_0x_1x_2x_3x_6x_7x_8x_9 \oplus
x_4x_6x_7x_8x_9 \oplus x_1x_4x_6x_7x_8x_9 \oplus x_0x_1x_4x_6x_7x_8x_9
\oplus x_0x_2x_4x_6x_7x_8x_9 \oplus x_1x_2x_4x_6x_7x_8x_9 \oplus
x_3x_4x_6x_7x_8x_9 \oplus x_0x_3x_4x_6x_7x_8x_9 \oplus
x_0x_1x_3x_4x_6x_7x_8x_9 \oplus x_2x_3x_4x_6x_7x_8x_9 \oplus
x_1x_2x_3x_4x_6x_7x_8x_9 \oplus x_0x_1x_2x_3x_4x_6x_7x_8x_9 \oplus
x_0x_5x_6x_7x_8x_9 \oplus x_1x_5x_6x_7x_8x_9 \oplus x_2x_5x_6x_7x_8x_9
\oplus x_0x_2x_5x_6x_7x_8x_9 \oplus x_0x_1x_2x_5x_6x_7x_8x_9 \oplus
x_3x_5x_6x_7x_8x_9 \oplus x_1x_3x_5x_6x_7x_8x_9 \oplus
x_0x_1x_3x_5x_6x_7x_8x_9 \oplus x_0x_2x_3x_5x_6x_7x_8x_9 \oplus
x_1x_2x_3x_5x_6x_7x_8x_9 \oplus x_4x_5x_6x_7x_8x_9 \oplus
x_0x_4x_5x_6x_7x_8x_9 \oplus x_0x_1x_4x_5x_6x_7x_8x_9 \oplus
x_2x_4x_5x_6x_7x_8x_9 \oplus x_1x_2x_4x_5x_6x_7x_8x_9 \oplus
x_0x_1x_2x_4x_5x_6x_7x_8x_9 \oplus x_0x_3x_4x_5x_6x_7x_8x_9 \oplus
x_1x_3x_4x_5x_6x_7x_8x_9 \oplus x_2x_3x_4x_5x_6x_7x_8x_9 \oplus
x_0x_2x_3x_4x_5x_6x_7x_8x_9 \oplus x_0x_1x_2x_3x_4x_5x_6x_7x_8x_9 = 0$

\bigskip
%%%%%%%%%%%%%%%%%%%%%%%%%%%%%%%%%%%%%%%%%%%%%%%%%%%%%%%%%%%%%%%%%%%%%%%%%%%%%%%%%%%%%%%%%%
\twocolumngrid
\bigskip
\bibliography{cites.bib}

%apsrev4-2.bst 2019-01-14 (MD) hand-edited version of apsrev4-1.bst
%Control: key (0)
%Control: author (8) initials jnrlst
%Control: editor formatted (1) identically to author
%Control: production of article title (0) allowed
%Control: page (0) single
%Control: year (1) truncated
%Control: production of eprint (0) enabled
\begin{thebibliography}{37}%
\makeatletter
\providecommand \@ifxundefined [1]{%
 \@ifx{#1\undefined}
}%
\providecommand \@ifnum [1]{%
 \ifnum #1\expandafter \@firstoftwo
 \else \expandafter \@secondoftwo
 \fi
}%
\providecommand \@ifx [1]{%
 \ifx #1\expandafter \@firstoftwo
 \else \expandafter \@secondoftwo
 \fi
}%
\providecommand \natexlab [1]{#1}%
\providecommand \enquote  [1]{``#1''}%
\providecommand \bibnamefont  [1]{#1}%
\providecommand \bibfnamefont [1]{#1}%
\providecommand \citenamefont [1]{#1}%
\providecommand \href@noop [0]{\@secondoftwo}%
\providecommand \href [0]{\begingroup \@sanitize@url \@href}%
\providecommand \@href[1]{\@@startlink{#1}\@@href}%
\providecommand \@@href[1]{\endgroup#1\@@endlink}%
\providecommand \@sanitize@url [0]{\catcode `\\12\catcode `\$12\catcode
  `\&12\catcode `\#12\catcode `\^12\catcode `\_12\catcode `\%12\relax}%
\providecommand \@@startlink[1]{}%
\providecommand \@@endlink[0]{}%
\providecommand \url  [0]{\begingroup\@sanitize@url \@url }%
\providecommand \@url [1]{\endgroup\@href {#1}{\urlprefix }}%
\providecommand \urlprefix  [0]{URL }%
\providecommand \Eprint [0]{\href }%
\providecommand \doibase [0]{https://doi.org/}%
\providecommand \selectlanguage [0]{\@gobble}%
\providecommand \bibinfo  [0]{\@secondoftwo}%
\providecommand \bibfield  [0]{\@secondoftwo}%
\providecommand \translation [1]{[#1]}%
\providecommand \BibitemOpen [0]{}%
\providecommand \bibitemStop [0]{}%
\providecommand \bibitemNoStop [0]{.\EOS\space}%
\providecommand \EOS [0]{\spacefactor3000\relax}%
\providecommand \BibitemShut  [1]{\csname bibitem#1\endcsname}%
\let\auto@bib@innerbib\@empty
%</preamble>
\bibitem [{\citenamefont {Barnett}\ \emph {et~al.}(2021)\citenamefont
  {Barnett}, \citenamefont {Jeffers},\ and\ \citenamefont
  {Pegg}}]{sym13040586}%
  \BibitemOpen
  \bibfield  {author} {\bibinfo {author} {\bibfnamefont {S.~M.}\ \bibnamefont
  {Barnett}}, \bibinfo {author} {\bibfnamefont {J.}~\bibnamefont {Jeffers}},\
  and\ \bibinfo {author} {\bibfnamefont {D.~T.}\ \bibnamefont {Pegg}},\
  }\bibfield  {title} {\bibinfo {title} {Quantum retrodiction: Foundations and
  controversies},\ }\bibfield  {journal} {\bibinfo  {journal} {Symmetry}\
  }\textbf {\bibinfo {volume} {13}},\ \href
  {https://doi.org/10.3390/sym13040586} {10.3390/sym13040586} (\bibinfo {year}
  {2021})\BibitemShut {NoStop}%
\bibitem [{\citenamefont {Aharonov}\ and\ \citenamefont
  {Vaidman}(2008)}]{Aharonov2008}%
  \BibitemOpen
  \bibfield  {author} {\bibinfo {author} {\bibfnamefont {Y.}~\bibnamefont
  {Aharonov}}\ and\ \bibinfo {author} {\bibfnamefont {L.}~\bibnamefont
  {Vaidman}},\ }\bibinfo {title} {The two-state vector formalism: An updated
  review},\ in\ \href {https://doi.org/10.1007/978-3-540-73473-4_13} {\emph
  {\bibinfo {booktitle} {Time in Quantum Mechanics}}},\ \bibinfo {editor}
  {edited by\ \bibinfo {editor} {\bibfnamefont {J.}~\bibnamefont {Muga}},
  \bibinfo {editor} {\bibfnamefont {R.~S.}\ \bibnamefont {Mayato}},\ and\
  \bibinfo {editor} {\bibfnamefont {{\'I}.}~\bibnamefont {Egusquiza}}}\
  (\bibinfo  {publisher} {Springer Berlin Heidelberg},\ \bibinfo {address}
  {Berlin, Heidelberg},\ \bibinfo {year} {2008})\ pp.\ \bibinfo {pages}
  {399--447}\BibitemShut {NoStop}%
\bibitem [{\citenamefont {Watanabe}(1955)}]{RevModPhys.27.179}%
  \BibitemOpen
  \bibfield  {author} {\bibinfo {author} {\bibfnamefont {S.}~\bibnamefont
  {Watanabe}},\ }\bibfield  {title} {\bibinfo {title} {Symmetry of physical
  laws. {Part III}. prediction and retrodiction},\ }\href
  {https://doi.org/10.1103/RevModPhys.27.179} {\bibfield  {journal} {\bibinfo
  {journal} {Rev. Mod. Phys.}\ }\textbf {\bibinfo {volume} {27}},\ \bibinfo
  {pages} {179} (\bibinfo {year} {1955})}\BibitemShut {NoStop}%
\bibitem [{\citenamefont {Futamura}(1983)}]{futamura}%
  \BibitemOpen
  \bibfield  {author} {\bibinfo {author} {\bibfnamefont {Y.}~\bibnamefont
  {Futamura}},\ }\bibfield  {title} {\bibinfo {title} {Partial computation of
  programs},\ }in\ \href@noop {} {\emph {\bibinfo {booktitle} {RIMS Symposia on
  Software Science and Engineering}}},\ \bibinfo {editor} {edited by\ \bibinfo
  {editor} {\bibfnamefont {E.}~\bibnamefont {Goto}}, \bibinfo {editor}
  {\bibfnamefont {K.}~\bibnamefont {Furukawa}}, \bibinfo {editor}
  {\bibfnamefont {R.}~\bibnamefont {Nakajima}}, \bibinfo {editor}
  {\bibfnamefont {I.}~\bibnamefont {Nakata}},\ and\ \bibinfo {editor}
  {\bibfnamefont {A.}~\bibnamefont {Yonezawa}}}\ (\bibinfo  {publisher}
  {Springer Berlin Heidelberg},\ \bibinfo {address} {Berlin, Heidelberg},\
  \bibinfo {year} {1983})\ pp.\ \bibinfo {pages} {1--35}\BibitemShut {NoStop}%
\bibitem [{\citenamefont {Bernstein}\ and\ \citenamefont
  {Vazirani}(1997)}]{doi:10.1137/S0097539796300921}%
  \BibitemOpen
  \bibfield  {author} {\bibinfo {author} {\bibfnamefont {E.}~\bibnamefont
  {Bernstein}}\ and\ \bibinfo {author} {\bibfnamefont {U.}~\bibnamefont
  {Vazirani}},\ }\bibfield  {title} {\bibinfo {title} {Quantum complexity
  theory},\ }\href {https://doi.org/10.1137/S0097539796300921} {\bibfield
  {journal} {\bibinfo  {journal} {SIAM Journal on Computing}\ }\textbf
  {\bibinfo {volume} {26}},\ \bibinfo {pages} {1411} (\bibinfo {year}
  {1997})},\ \Eprint
  {https://arxiv.org/abs/https://doi.org/10.1137/S0097539796300921}
  {https://doi.org/10.1137/S0097539796300921} \BibitemShut {NoStop}%
\bibitem [{\citenamefont {Deutsch}(1985)}]{deutsch}%
  \BibitemOpen
  \bibfield  {author} {\bibinfo {author} {\bibfnamefont {D.}~\bibnamefont
  {Deutsch}},\ }\bibfield  {title} {\bibinfo {title} {Quantum theory, the
  {C}hurch–{T}uring principle and the universal quantum computer},\
  }\href@noop {} {\bibfield  {journal} {\bibinfo  {journal} {Proc. R. Soc.
  Lond. A 400}\ } (\bibinfo {year} {1985})}\BibitemShut {NoStop}%
\bibitem [{\citenamefont {Deutsch}\ and\ \citenamefont
  {Jozsa}(1992)}]{deutschJozsa}%
  \BibitemOpen
  \bibfield  {author} {\bibinfo {author} {\bibfnamefont {D.}~\bibnamefont
  {Deutsch}}\ and\ \bibinfo {author} {\bibfnamefont {R.}~\bibnamefont
  {Jozsa}},\ }\bibfield  {title} {\bibinfo {title} {Rapid solution of problems
  by quantum computation},\ }\href@noop {} {\bibfield  {journal} {\bibinfo
  {journal} {Proc. R. Soc. Lond. A 439}\ } (\bibinfo {year}
  {1992})}\BibitemShut {NoStop}%
\bibitem [{\citenamefont {Simon}(1994)}]{365701}%
  \BibitemOpen
  \bibfield  {author} {\bibinfo {author} {\bibfnamefont {D.}~\bibnamefont
  {Simon}},\ }\bibfield  {title} {\bibinfo {title} {On the power of quantum
  computation},\ }in\ \href {https://doi.org/10.1109/SFCS.1994.365701} {\emph
  {\bibinfo {booktitle} {Proceedings 35th Annual Symposium on Foundations of
  Computer Science}}}\ (\bibinfo {year} {1994})\ pp.\ \bibinfo {pages}
  {116--123}\BibitemShut {NoStop}%
\bibitem [{\citenamefont {Shor}(1997)}]{doi:10.1137/S0097539795293172}%
  \BibitemOpen
  \bibfield  {author} {\bibinfo {author} {\bibfnamefont {P.~W.}\ \bibnamefont
  {Shor}},\ }\bibfield  {title} {\bibinfo {title} {Polynomial-time algorithms
  for prime factorization and discrete logarithms on a quantum computer},\
  }\href {https://doi.org/10.1137/S0097539795293172} {\bibfield  {journal}
  {\bibinfo  {journal} {SIAM Journal on Computing}\ }\textbf {\bibinfo {volume}
  {26}},\ \bibinfo {pages} {1484} (\bibinfo {year} {1997})},\ \Eprint
  {https://arxiv.org/abs/https://doi.org/10.1137/S0097539795293172}
  {https://doi.org/10.1137/S0097539795293172} \BibitemShut {NoStop}%
\bibitem [{\citenamefont {Mermin}(2007)}]{mermin_2007}%
  \BibitemOpen
  \bibfield  {author} {\bibinfo {author} {\bibfnamefont {N.~D.}\ \bibnamefont
  {Mermin}},\ }\href {https://doi.org/10.1017/CBO9780511813870} {\emph
  {\bibinfo {title} {Quantum Computer Science: An Introduction}}}\ (\bibinfo
  {publisher} {Cambridge University Press},\ \bibinfo {year}
  {2007})\BibitemShut {NoStop}%
\bibitem [{\citenamefont {Nielsen}\ and\ \citenamefont
  {Chuang}(2010)}]{nielsen_chuang_2010}%
  \BibitemOpen
  \bibfield  {author} {\bibinfo {author} {\bibfnamefont {M.~A.}\ \bibnamefont
  {Nielsen}}\ and\ \bibinfo {author} {\bibfnamefont {I.~L.}\ \bibnamefont
  {Chuang}},\ }\href {https://doi.org/10.1017/CBO9780511976667} {\emph
  {\bibinfo {title} {Quantum Computation and Quantum Information: 10th
  Anniversary Edition}}}\ (\bibinfo  {publisher} {Cambridge University Press},\
  \bibinfo {year} {2010})\BibitemShut {NoStop}%
\bibitem [{\citenamefont {Grover}(1996)}]{10.1145/237814.237866}%
  \BibitemOpen
  \bibfield  {author} {\bibinfo {author} {\bibfnamefont {L.~K.}\ \bibnamefont
  {Grover}},\ }\bibfield  {title} {\bibinfo {title} {A fast quantum mechanical
  algorithm for database search},\ }in\ \href
  {https://doi.org/10.1145/237814.237866} {\emph {\bibinfo {booktitle}
  {Proceedings of the Twenty-Eighth Annual ACM Symposium on Theory of
  Computing}}},\ \bibinfo {series and number} {STOC '96}\ (\bibinfo
  {publisher} {Association for Computing Machinery},\ \bibinfo {address} {New
  York, NY, USA},\ \bibinfo {year} {1996})\ p.\ \bibinfo {pages}
  {212–219}\BibitemShut {NoStop}%
\bibitem [{\citenamefont {Komargodski}\ \emph {et~al.}(2019)\citenamefont
  {Komargodski}, \citenamefont {Naor},\ and\ \citenamefont
  {Yogev}}]{10.1145/3341106}%
  \BibitemOpen
  \bibfield  {author} {\bibinfo {author} {\bibfnamefont {I.}~\bibnamefont
  {Komargodski}}, \bibinfo {author} {\bibfnamefont {M.}~\bibnamefont {Naor}},\
  and\ \bibinfo {author} {\bibfnamefont {E.}~\bibnamefont {Yogev}},\ }\bibfield
   {title} {\bibinfo {title} {White-box vs. black-box complexity of search
  problems: Ramsey and graph property testing},\ }\bibfield  {journal}
  {\bibinfo  {journal} {J. ACM}\ }\textbf {\bibinfo {volume} {66}},\ \href
  {https://doi.org/10.1145/3341106} {10.1145/3341106} (\bibinfo {year}
  {2019})\BibitemShut {NoStop}%
\bibitem [{\citenamefont {Burnett}\ \emph {et~al.}(2004)\citenamefont
  {Burnett}, \citenamefont {Millan}, \citenamefont {Dawson},\ and\
  \citenamefont {Clark}}]{quteprints21763}%
  \BibitemOpen
  \bibfield  {author} {\bibinfo {author} {\bibfnamefont {L.}~\bibnamefont
  {Burnett}}, \bibinfo {author} {\bibfnamefont {W.}~\bibnamefont {Millan}},
  \bibinfo {author} {\bibfnamefont {E.}~\bibnamefont {Dawson}},\ and\ \bibinfo
  {author} {\bibfnamefont {A.}~\bibnamefont {Clark}},\ }\bibfield  {title}
  {\bibinfo {title} {Simpler methods for generating better boolean functions
  with good cryptographic properties},\ }\href
  {https://eprints.qut.edu.au/21763/} {\bibfield  {journal} {\bibinfo
  {journal} {Australasian Journal of Combinatorics}\ }\textbf {\bibinfo
  {volume} {29}},\ \bibinfo {pages} {231} (\bibinfo {year} {2004})}\BibitemShut
  {NoStop}%
\bibitem [{\citenamefont {Abbott}(2012)}]{djdeq}%
  \BibitemOpen
  \bibfield  {author} {\bibinfo {author} {\bibfnamefont {A.~A.}\ \bibnamefont
  {Abbott}},\ }\bibfield  {title} {\bibinfo {title} {The {Deutsch-Jozsa}
  problem: de-quantization and entanglement},\ }\href@noop {} {\bibfield
  {journal} {\bibinfo  {journal} {Natural Computing}\ }\textbf {\bibinfo
  {volume} {11}} (\bibinfo {year} {2012})}\BibitemShut {NoStop}%
\bibitem [{\citenamefont {Trakhtenbrot}(1984)}]{4640789}%
  \BibitemOpen
  \bibfield  {author} {\bibinfo {author} {\bibfnamefont {B.}~\bibnamefont
  {Trakhtenbrot}},\ }\bibfield  {title} {\bibinfo {title} {A survey of
  {R}ussian approaches to {P}erebor (brute-force searches) algorithms},\ }\href
  {https://doi.org/10.1109/MAHC.1984.10036} {\bibfield  {journal} {\bibinfo
  {journal} {Annals of the History of Computing}\ }\textbf {\bibinfo {volume}
  {6}},\ \bibinfo {pages} {384} (\bibinfo {year} {1984})}\BibitemShut {NoStop}%
\bibitem [{\citenamefont {Karp}(1972)}]{Karp1972}%
  \BibitemOpen
  \bibfield  {author} {\bibinfo {author} {\bibfnamefont {R.~M.}\ \bibnamefont
  {Karp}},\ }\bibinfo {title} {Reducibility among combinatorial problems},\ in\
  \href {https://doi.org/10.1007/978-1-4684-2001-2_9} {\emph {\bibinfo
  {booktitle} {Complexity of Computer Computations: Proceedings of a symposium
  on the Complexity of Computer Computations, held March 20--22, 1972, at the
  IBM Thomas J. Watson Research Center, Yorktown Heights, New York, and
  sponsored by the Office of Naval Research, Mathematics Program, IBM World
  Trade Corporation, and the IBM Research Mathematical Sciences Department}}},\
  \bibinfo {editor} {edited by\ \bibinfo {editor} {\bibfnamefont {R.~E.}\
  \bibnamefont {Miller}}, \bibinfo {editor} {\bibfnamefont {J.~W.}\
  \bibnamefont {Thatcher}},\ and\ \bibinfo {editor} {\bibfnamefont {J.~D.}\
  \bibnamefont {Bohlinger}}}\ (\bibinfo  {publisher} {Springer US},\ \bibinfo
  {address} {Boston, MA},\ \bibinfo {year} {1972})\ pp.\ \bibinfo {pages}
  {85--103}\BibitemShut {NoStop}%
\bibitem [{\citenamefont {Cook}(1971)}]{10.1145/800157.805047}%
  \BibitemOpen
  \bibfield  {author} {\bibinfo {author} {\bibfnamefont {S.~A.}\ \bibnamefont
  {Cook}},\ }\bibfield  {title} {\bibinfo {title} {The complexity of
  theorem-proving procedures},\ }in\ \href
  {https://doi.org/10.1145/800157.805047} {\emph {\bibinfo {booktitle}
  {Proceedings of the Third Annual ACM Symposium on Theory of Computing}}},\
  \bibinfo {series and number} {STOC '71}\ (\bibinfo  {publisher} {Association
  for Computing Machinery},\ \bibinfo {address} {New York, NY, USA},\ \bibinfo
  {year} {1971})\ p.\ \bibinfo {pages} {151–158}\BibitemShut {NoStop}%
\bibitem [{\citenamefont {Bocharov}\ \emph {et~al.}(2016)\citenamefont
  {Bocharov}, \citenamefont {Cui}, \citenamefont {Roetteler},\ and\
  \citenamefont {Svore}}]{10.5555/3179473.3179481}%
  \BibitemOpen
  \bibfield  {author} {\bibinfo {author} {\bibfnamefont {A.}~\bibnamefont
  {Bocharov}}, \bibinfo {author} {\bibfnamefont {S.~X.}\ \bibnamefont {Cui}},
  \bibinfo {author} {\bibfnamefont {M.}~\bibnamefont {Roetteler}},\ and\
  \bibinfo {author} {\bibfnamefont {K.~M.}\ \bibnamefont {Svore}},\ }\bibfield
  {title} {\bibinfo {title} {Improved quantum ternary arithmetic},\ }\href@noop
  {} {\bibfield  {journal} {\bibinfo  {journal} {Quantum Info. Comput.}\
  }\textbf {\bibinfo {volume} {16}},\ \bibinfo {pages} {862–884} (\bibinfo
  {year} {2016})}\BibitemShut {NoStop}%
\bibitem [{\citenamefont {Vedral}\ \emph {et~al.}(1996)\citenamefont {Vedral},
  \citenamefont {Barenco},\ and\ \citenamefont {Ekert}}]{PhysRevA.54.147}%
  \BibitemOpen
  \bibfield  {author} {\bibinfo {author} {\bibfnamefont {V.}~\bibnamefont
  {Vedral}}, \bibinfo {author} {\bibfnamefont {A.}~\bibnamefont {Barenco}},\
  and\ \bibinfo {author} {\bibfnamefont {A.}~\bibnamefont {Ekert}},\ }\bibfield
   {title} {\bibinfo {title} {Quantum networks for elementary arithmetic
  operations},\ }\href {https://doi.org/10.1103/PhysRevA.54.147} {\bibfield
  {journal} {\bibinfo  {journal} {Phys. Rev. A}\ }\textbf {\bibinfo {volume}
  {54}},\ \bibinfo {pages} {147} (\bibinfo {year} {1996})}\BibitemShut
  {NoStop}%
\bibitem [{\citenamefont {Geller}\ and\ \citenamefont
  {Zhou}(2013)}]{shorFermat}%
  \BibitemOpen
  \bibfield  {author} {\bibinfo {author} {\bibfnamefont {M.~R.}\ \bibnamefont
  {Geller}}\ and\ \bibinfo {author} {\bibfnamefont {Z.}~\bibnamefont {Zhou}},\
  }\bibfield  {title} {\bibinfo {title} {Factoring 51 and 85 with 8 qubits},\
  }\href@noop {} {\bibfield  {journal} {\bibinfo  {journal} {Scientific Reports
  (3023)}\ }\textbf {\bibinfo {volume} {3}} (\bibinfo {year}
  {2013})}\BibitemShut {NoStop}%
\bibitem [{\citenamefont {Jozsa}\ and\ \citenamefont
  {Linden}(2003)}]{10.2307/3560059}%
  \BibitemOpen
  \bibfield  {author} {\bibinfo {author} {\bibfnamefont {R.}~\bibnamefont
  {Jozsa}}\ and\ \bibinfo {author} {\bibfnamefont {N.}~\bibnamefont {Linden}},\
  }\bibfield  {title} {\bibinfo {title} {On the role of entanglement in
  quantum-computational speed-up},\ }\href
  {http://www.jstor.org/stable/3560059} {\bibfield  {journal} {\bibinfo
  {journal} {Proceedings: Mathematical, Physical and Engineering Sciences}\
  }\textbf {\bibinfo {volume} {459}},\ \bibinfo {pages} {2011} (\bibinfo {year}
  {2003})}\BibitemShut {NoStop}%
\bibitem [{\citenamefont {Barnum}\ \emph {et~al.}(2004)\citenamefont {Barnum},
  \citenamefont {Knill}, \citenamefont {Ortiz}, \citenamefont {Somma},\ and\
  \citenamefont {Viola}}]{GE2004}%
  \BibitemOpen
  \bibfield  {author} {\bibinfo {author} {\bibfnamefont {H.}~\bibnamefont
  {Barnum}}, \bibinfo {author} {\bibfnamefont {E.}~\bibnamefont {Knill}},
  \bibinfo {author} {\bibfnamefont {G.}~\bibnamefont {Ortiz}}, \bibinfo
  {author} {\bibfnamefont {R.}~\bibnamefont {Somma}},\ and\ \bibinfo {author}
  {\bibfnamefont {L.}~\bibnamefont {Viola}},\ }\bibfield  {title} {\bibinfo
  {title} {A subsystem-independent generalization of entanglement},\ }\href
  {https://doi.org/10.1103/PhysRevLett.92.107902} {\bibfield  {journal}
  {\bibinfo  {journal} {Phys. Rev. Lett.}\ }\textbf {\bibinfo {volume} {92}},\
  \bibinfo {pages} {107902} (\bibinfo {year} {2004})}\BibitemShut {NoStop}%
\bibitem [{\citenamefont {Aharonov}(2003)}]{haduniv}%
  \BibitemOpen
  \bibfield  {author} {\bibinfo {author} {\bibfnamefont {D.}~\bibnamefont
  {Aharonov}},\ }\bibfield  {title} {\bibinfo {title} {A simple proof that
  {T}offoli and {H}adamard are quantum universal},\ }\href@noop {} {\bibfield
  {journal} {\bibinfo  {journal} {arXiv:quant-ph/0301040}\ } (\bibinfo {year}
  {2003})}\BibitemShut {NoStop}%
\bibitem [{\citenamefont {Boyer}\ \emph {et~al.}(1975)\citenamefont {Boyer},
  \citenamefont {Elspas},\ and\ \citenamefont
  {Levitt}}]{10.1145/390016.808445}%
  \BibitemOpen
  \bibfield  {author} {\bibinfo {author} {\bibfnamefont {R.~S.}\ \bibnamefont
  {Boyer}}, \bibinfo {author} {\bibfnamefont {B.}~\bibnamefont {Elspas}},\ and\
  \bibinfo {author} {\bibfnamefont {K.~N.}\ \bibnamefont {Levitt}},\ }\bibfield
   {title} {\bibinfo {title} {Select—a formal system for testing and
  debugging programs by symbolic execution},\ }\href
  {https://doi.org/10.1145/390016.808445} {\bibfield  {journal} {\bibinfo
  {journal} {SIGPLAN Not.}\ }\textbf {\bibinfo {volume} {10}},\ \bibinfo
  {pages} {234–245} (\bibinfo {year} {1975})}\BibitemShut {NoStop}%
\bibitem [{\citenamefont {King}(1976)}]{10.1145/360248.360252}%
  \BibitemOpen
  \bibfield  {author} {\bibinfo {author} {\bibfnamefont {J.~C.}\ \bibnamefont
  {King}},\ }\bibfield  {title} {\bibinfo {title} {Symbolic execution and
  program testing},\ }\href {https://doi.org/10.1145/360248.360252} {\bibfield
  {journal} {\bibinfo  {journal} {Commun. ACM}\ }\textbf {\bibinfo {volume}
  {19}},\ \bibinfo {pages} {385–394} (\bibinfo {year} {1976})}\BibitemShut
  {NoStop}%
\bibitem [{\citenamefont {Howden}(1976)}]{howden}%
  \BibitemOpen
  \bibfield  {author} {\bibinfo {author} {\bibfnamefont {W.~E.}\ \bibnamefont
  {Howden}},\ }\bibfield  {title} {\bibinfo {title} {Experiments with a
  symbolic evaluation system},\ }in\ \href@noop {} {\emph {\bibinfo {booktitle}
  {Proceedings of the National Computer Conference}}}\ (\bibinfo {year}
  {1976})\BibitemShut {NoStop}%
\bibitem [{\citenamefont {Clarke}(1976)}]{10.1145/800191.805647}%
  \BibitemOpen
  \bibfield  {author} {\bibinfo {author} {\bibfnamefont {L.~A.}\ \bibnamefont
  {Clarke}},\ }\bibfield  {title} {\bibinfo {title} {A program testing
  system},\ }in\ \href {https://doi.org/10.1145/800191.805647} {\emph {\bibinfo
  {booktitle} {Proceedings of the 1976 Annual Conference}}},\ \bibinfo {series
  and number} {ACM '76}\ (\bibinfo  {publisher} {Association for Computing
  Machinery},\ \bibinfo {address} {New York, NY, USA},\ \bibinfo {year}
  {1976})\ p.\ \bibinfo {pages} {488–491}\BibitemShut {NoStop}%
\bibitem [{\citenamefont {Baldoni}\ \emph {et~al.}(2018)\citenamefont
  {Baldoni}, \citenamefont {Coppa}, \citenamefont {D’elia}, \citenamefont
  {Demetrescu},\ and\ \citenamefont {Finocchi}}]{10.1145/3182657}%
  \BibitemOpen
  \bibfield  {author} {\bibinfo {author} {\bibfnamefont {R.}~\bibnamefont
  {Baldoni}}, \bibinfo {author} {\bibfnamefont {E.}~\bibnamefont {Coppa}},
  \bibinfo {author} {\bibfnamefont {D.~C.}\ \bibnamefont {D’elia}}, \bibinfo
  {author} {\bibfnamefont {C.}~\bibnamefont {Demetrescu}},\ and\ \bibinfo
  {author} {\bibfnamefont {I.}~\bibnamefont {Finocchi}},\ }\bibfield  {title}
  {\bibinfo {title} {A survey of symbolic execution techniques},\ }\bibfield
  {journal} {\bibinfo  {journal} {ACM Comput. Surv.}\ }\textbf {\bibinfo
  {volume} {51}},\ \href {https://doi.org/10.1145/3182657} {10.1145/3182657}
  (\bibinfo {year} {2018})\BibitemShut {NoStop}%
\bibitem [{\citenamefont {Wegener}(1987)}]{10.5555/35517}%
  \BibitemOpen
  \bibfield  {author} {\bibinfo {author} {\bibfnamefont {I.}~\bibnamefont
  {Wegener}},\ }\href@noop {} {\emph {\bibinfo {title} {The Complexity of
  Boolean Functions}}}\ (\bibinfo  {publisher} {John Wiley \& Sons, Inc.},\
  \bibinfo {address} {USA},\ \bibinfo {year} {1987})\BibitemShut {NoStop}%
\bibitem [{\citenamefont {Tokareva}(2015)}]{TOKAREVA20151}%
  \BibitemOpen
  \bibfield  {author} {\bibinfo {author} {\bibfnamefont {N.}~\bibnamefont
  {Tokareva}},\ }\bibfield  {title} {\bibinfo {title} {Chapter 1 - {Boolean}
  functions},\ }in\ \href
  {https://doi.org/https://doi.org/10.1016/B978-0-12-802318-1.00001-7} {\emph
  {\bibinfo {booktitle} {Bent Functions}}},\ \bibinfo {editor} {edited by\
  \bibinfo {editor} {\bibfnamefont {N.}~\bibnamefont {Tokareva}}}\ (\bibinfo
  {publisher} {Academic Press},\ \bibinfo {address} {Boston},\ \bibinfo {year}
  {2015})\ pp.\ \bibinfo {pages} {1--15}\BibitemShut {NoStop}%
\bibitem [{\citenamefont {Gottesman}(1998)}]{GKThm}%
  \BibitemOpen
  \bibfield  {author} {\bibinfo {author} {\bibfnamefont {D.}~\bibnamefont
  {Gottesman}},\ }\bibfield  {title} {\bibinfo {title} {The {H}eisenberg
  representation of quantum computers},\ }\href@noop {} {\bibfield  {journal}
  {\bibinfo  {journal} {arXiv:quant-ph/9807006}\ } (\bibinfo {year}
  {1998})}\BibitemShut {NoStop}%
\bibitem [{has(2010)}]{haskell}%
  \BibitemOpen
  \bibfield  {title} {\bibinfo {title} {Haskell 2010 language report},\
  }\href@noop {} {\bibfield  {journal} {\bibinfo  {journal} {Available online
  \url{http://www. haskell. org/}}\ } (\bibinfo {year} {2010})}\BibitemShut
  {NoStop}%
\bibitem [{\citenamefont {Rogaway}\ and\ \citenamefont
  {Shrimpton}(2004)}]{10.1007/978-3-540-25937-4_24}%
  \BibitemOpen
  \bibfield  {author} {\bibinfo {author} {\bibfnamefont {P.}~\bibnamefont
  {Rogaway}}\ and\ \bibinfo {author} {\bibfnamefont {T.}~\bibnamefont
  {Shrimpton}},\ }\bibfield  {title} {\bibinfo {title} {Cryptographic
  hash-function basics: Definitions, implications, and separations for preimage
  resistance, second-preimage resistance, and collision resistance},\ }in\
  \href@noop {} {\emph {\bibinfo {booktitle} {Fast Software Encryption}}},\
  \bibinfo {editor} {edited by\ \bibinfo {editor} {\bibfnamefont
  {B.}~\bibnamefont {Roy}}\ and\ \bibinfo {editor} {\bibfnamefont
  {W.}~\bibnamefont {Meier}}}\ (\bibinfo  {publisher} {Springer Berlin
  Heidelberg},\ \bibinfo {address} {Berlin, Heidelberg},\ \bibinfo {year}
  {2004})\ pp.\ \bibinfo {pages} {371--388}\BibitemShut {NoStop}%
\bibitem [{\citenamefont {Klotz}\ \emph {et~al.}(2013)\citenamefont {Klotz},
  \citenamefont {Bossert},\ and\ \citenamefont {Schober}}]{Klotz2013}%
  \BibitemOpen
  \bibfield  {author} {\bibinfo {author} {\bibfnamefont {J.~G.}\ \bibnamefont
  {Klotz}}, \bibinfo {author} {\bibfnamefont {M.}~\bibnamefont {Bossert}},\
  and\ \bibinfo {author} {\bibfnamefont {S.}~\bibnamefont {Schober}},\
  }\bibfield  {title} {\bibinfo {title} {Computing preimages of boolean
  networks},\ }\href {https://doi.org/10.1186/1471-2105-14-S10-S4} {\bibfield
  {journal} {\bibinfo  {journal} {BMC Bioinformatics}\ }\textbf {\bibinfo
  {volume} {14}},\ \bibinfo {pages} {S4} (\bibinfo {year} {2013})}\BibitemShut
  {NoStop}%
\bibitem [{\citenamefont {Akutsu}\ \emph {et~al.}(2009)\citenamefont {Akutsu},
  \citenamefont {Hayashida}, \citenamefont {Zhang}, \citenamefont {Ching},\
  and\ \citenamefont {Ng}}]{akutsu2009analyses}%
  \BibitemOpen
  \bibfield  {author} {\bibinfo {author} {\bibfnamefont {T.}~\bibnamefont
  {Akutsu}}, \bibinfo {author} {\bibfnamefont {M.}~\bibnamefont {Hayashida}},
  \bibinfo {author} {\bibfnamefont {S.-Q.}\ \bibnamefont {Zhang}}, \bibinfo
  {author} {\bibfnamefont {W.-K.}\ \bibnamefont {Ching}},\ and\ \bibinfo
  {author} {\bibfnamefont {M.~K.}\ \bibnamefont {Ng}},\ }\bibfield  {title}
  {\bibinfo {title} {Analyses and algorithms for predecessor and control
  problems for boolean networks of bounded indegree},\ }\href@noop {}
  {\bibfield  {journal} {\bibinfo  {journal} {Information and Media
  Technologies}\ }\textbf {\bibinfo {volume} {4}},\ \bibinfo {pages} {338}
  (\bibinfo {year} {2009})}\BibitemShut {NoStop}%
\bibitem [{\citenamefont {Kwok}\ and\ \citenamefont {Tsang}(2004)}]{1353287}%
  \BibitemOpen
  \bibfield  {author} {\bibinfo {author} {\bibfnamefont {J.-Y.}\ \bibnamefont
  {Kwok}}\ and\ \bibinfo {author} {\bibfnamefont {I.-H.}\ \bibnamefont
  {Tsang}},\ }\bibfield  {title} {\bibinfo {title} {The pre-image problem in
  kernel methods},\ }\href {https://doi.org/10.1109/TNN.2004.837781} {\bibfield
   {journal} {\bibinfo  {journal} {IEEE Transactions on Neural Networks}\
  }\textbf {\bibinfo {volume} {15}},\ \bibinfo {pages} {1517} (\bibinfo {year}
  {2004})}\BibitemShut {NoStop}%
\end{thebibliography}%


%apsrev4-2.bst 2019-01-14 (MD) hand-edited version of apsrev4-1.bst
%Control: key (0)
%Control: author (8) initials jnrlst
%Control: editor formatted (1) identically to author
%Control: production of article title (0) allowed
%Control: page (0) single
%Control: year (1) truncated
%Control: production of eprint (0) enabled
%
\end{document}